\documentclass[preprint,hyper]{JHEP3} % 10pt is ignored!

\JHEPspecialurl{http://jhep.sissa.it/JOURNAL/JHEP3.tar.gz}

\usepackage{epsfig,multicol}

\def\s{\sigma}
\def\g{\gamma}

\def\o{\omega}
\def\tn{\textnormal}
\def\tr{\textnormal{Tr}}
\def\be{\begin{equation}}
\def\ee{\end{equation}}
\def\bea{\begin{eqnarray}}
\def\eea{\end{eqnarray}}
\newcommand\fverb{\setbox\pippobox=\hbox\bgroup\verb}
\newcommand\fverbdo{\egroup\medskip\noindent%
                        \fbox{\unhbox\pippobox}\ }
\newcommand\fverbit{\egroup\item[\fbox{\unhbox\pippobox}]}
\newbox\pippobox
\title{The general Leigh-Strassler deformation and Integrability}

\author{Daniel Bundzik \\
        School of Technology and Society, Malm\"{o} University, \\ \"{O}stra Varvsgatan 11A, S-205 06
        Malm\"{o}, Sweden
        \\  Department of Theoretical Physics, Lund
        University,\\ S\"{o}lvegatan 14A, S-223 62, Sweden \\
        E-mail: \email{Daniel.Bundzik@ts.mah.se}}
\author{Teresia M\aa nsson\\
        NORDITA, Blegdamsvej 17, DK-2100 Copenhagen, Denmark\\
        E-mail:  \email{teresia@nordita.dk}}
\preprint{\hepth{0512093}\\NORDITA-2005-76\\LU TP 05-46}

\abstract{ The success of the identification of the planar dilatation
operator of $\mathcal{N}=4$ SYM with an integrable spin chain
Hamiltonian has raised the question if this also is valid for a
deformed theory. Several deformations of SYM have recently been
under investigation in this context. In this work we consider the
general Leigh-Strassler deformation. For the generic case the
S-matrix techniques cannot be used to prove integrability. Instead
we use R-matrix techniques to study integrability. Some new
integrable points in the parameter space are found.}


\begin{document}
\section{Introduction}
In the last few years, several  new discoveries have shed light on
the AdS/CFT correspondence
\cite{Maldacena:1998re,Gubser:1998bc,Witten:1998qj}. This
correspondence maps strings moving in an $AdS_5\times S^5$
background to an $\mathcal{N}=4$
 supersymmetric Yang-Mills (SYM) theory. The eigenvalues
of the dilatation operator are mapped to the energies of  closed
string states \cite{Berenstein:2002jq}. A step in understanding  this duality better was the
discovery that the  dilatation operator of the $\mathcal{N}=4$  SYM
is proportional to the Hamiltonian of
 an integrable spin chain \cite{Minahan:2002ve,Beisert:2003yb,Beisert:2003tq}.

Recently, progress has been made to extend the
gauge/gravity-correspondence, in context of spin chains, towards
more realistic models with less supersymmetry
\cite{Frolov:2005ty,Frolov:2005iq,Beisert:2005he,Berenstein:2005ek,Ideguchi:2004wm}.
For instance, if the background geometry for the string is
$AdS_5\times W$,
 where $W$ is some compact manifold, the dual gauge theory
should still be conformal. Other geometries, mainly orbifolds of
$AdS_5\times S^5$, corresponding to non-conformal theories have also
been investigated \cite{Wang:2003cu,DeWolfe:2004zt}.

The success in using spin chains to study the duality beyond the BMN
limit motivates studies of integrability of deformed correlators.
One question that naturally arises in this context
 is whether integrability is related to supersymmetry, conformal
invariance or have more geometrical reasons.

The Leigh-Strassler deformations \cite{Leigh:1995ep}
 preserve $\mathcal{N}=1$ supersymmetry
and conformal invariance, at least up to one loop.
It is hence of great interest to investigate if there exist points
in the parameter space where the dilatation operator is mapped to an
integrable spin-chain Hamiltonian. This question has been under
investigation in \cite{Roiban:2003dw,
Berenstein:2004ys,Beisert:2005if,Freyhult:2005ws}. In
\cite{Berenstein:2004ys}
 this deformation was studied in a special case corresponding to
a $q$-deformed (often called $\beta$-deformed) commutator. It was
found that for the sector with three chiral fields the dilatation
operator is integrable for $q$ equals root of unity.

In reference \cite{Lunin:2005jy}, a way of generating supergravity
duals to the $\beta$-deformed field theory was introduced, and in
\cite{Frolov:2005ty,Frolov:2005iq,Kuzenko:2005gy} agreement
 between the supergravity sigma model and the coherent state action
coming from the spin chain describing the $\beta$-deformed
dilatation operator was demonstrated. This way of creating
supergravity duals was used in \cite{Frolov:2005dj} to construct a
three-parameter generalization of the $\beta$-deformed theory. The
gauge theory dual to this supergravity solution was found in
\cite{Frolov:2005dj,Beisert:2005if} for $q=e^{i\g_j}$ with $\g_j$
real, corresponding to certain phase deformations in the Lagrangian.
This gauge theory is referred to as twisted SYM, from which the
$\beta$-deformed theory is obtained when all the $\g_j=\beta$. The
result is that the theory is integrable for
 any $q=e^{i\gamma_j}$ with $\gamma_j$ real \cite{Beisert:2005if}. The
general case with complex $\g_j$ is not integrable
\cite{Berenstein:2004ys,Freyhult:2005ws}.

In the present work, the $q$-deformed analysis is extended to the
more general Leigh-Strassler deformations with an extra complex
parameter $h$, in order to find new integrable theories. A site
dependent transformation is found which relates the
$\gamma_j$-deformed case to a site dependent spin-chain
 Hamiltonian with nearest-neighbour interactions. In particular when all $\g_j$ are equal, the
transformation relates the $q$-deformed theory to the $h$-deformed
theory, {\it i.e.} the theory only involving the parameter $h$. In particular,
we find a new R-matrix, at least in the context of $\mathcal{N}=4$ SYM,
  for $q=0$ and $h=e^{i\theta}$ with $\theta$
real. We also find all R-matrices with
a linear dependence on the spectral parameter which give the
dilatation operator. A general ansatz for
the R-matrix is given. Unfortunately, the most general solution is
not found. However, we find a new hyperbolic R-matrix which
corresponds to a basis-transformed Hamiltonian with only diagonal
entries \cite{Freyhult:2005ws}.
 A reformulation of the general R-matrix shows that the structure
of the equations obtained from the
Yang-Baxter equations resemble the  equations obtained in the
eight vertex model. This gives a clear hint how to proceed.

In the dual supergravity theory, some attempts to construct
backgrounds for non-zero $h$ have been done
\cite{Aharony:2002hx,Fayyazuddin:2002vh}.
 Apart from the five-flux there is also a three-flux. A step
going beyond supergravity was taken in \cite{Niarchos:2002fc} where
the BMN limit was considered. We hope our results will make it
easier to find the supergravity dual of the general Leigh-Strassler
deformed theory.

%%%%%%%%%%%%%%%%%%%%%%%%%%%%%%%%%%%%%%%%%%%%%%%%%%%%%%%%%%%%%%%%%%%%%%%%%%
%  Section 2: Marginal deformations of N=4 supersymmetric Yang-Mills
%%%%%%%%%%%%%%%%%%%%%%%%%%%%%%%%%%%%%%%%%%%%%%%%%%%%%%%%%%%%%%%%%%%%%%%%%%

\section{Marginal deformations of $\mathcal{N}$=4 supersymmetric Yang-Mills}
To study marginal deformations of $\mathcal{N}=4$ SYM with $SU(N)$
gauge group, it is convenient to use $\mathcal{N}=1$ SYM
superfields. The six real scalar fields of the $\mathcal{N}=4$
vector multiplet are combined into the lowest order terms of three
complex $\mathcal{N}=1$ chiral superfields $\Phi_{0}$, $\Phi_{1}$
and $\Phi_{2}$.
It is well known that the $\mathcal{N}=1$
 superpotential
\be
W_{\mathcal{N}=1}=\frac{1}{3!}C_{abc}^{IJK}\Phi^a_{I}\Phi^b_{J}\Phi^c_{K}\,
, \label{superpotential} \ee where $C_{abc}^{IJK}$ is the coupling
constant, describes a finite theory at one-loop if the following two
conditions are fulfilled \cite{Parkes:1984dh,Jones:1984cx} \be
\label{NVZ} 3C_{2} (G) =\sum_{I} T(A_{I}),\hspace{1cm} \tn{and}
\hspace{1cm} C_{acd}^{IKL}\bar{C}^{bcd}_{JKL}=2g^2
T(A_{I})\delta_{a}^{\ b}\delta^{I}_{\ J}.\label{finite} \ee The
constant $C_{2}(G)$ is the quadratic Casimir operator defined
here\footnote{Our conventions are: $T^a$ are the $SU(N)$ group
generators, satisfying $T^{a}=T^{a\dag}$. The normalization of $T^a$
is given by $\tr(T^{a}T^{b})=\delta^{ab}/2$ from where it follows
that $\tr(T_{A}^{a}T_{A}^{b})=N\delta^{ab}$ in the adjoint
representation $A$.} as $C_{2}(G)\cdot \mathbf{1}=\delta_{ab}
T^{a}_{A}T^{b}_{A}$ where $A$ is the adjoint representation of the
group $G$ which in the present context is the symmetry group
$SU(N)$. The constant $T(M)$ is defined through $T(M)\delta^{ab}=\tr
(T^a_MT^b_M)$ for the representation $M$.
 The first condition of (\ref{NVZ}) implies that the
$\beta$-function is zero. For an $SU(N)$ group with the
superpotential (\ref{superpotential}) this is automatically
fulfilled. The choice $C_{abc}^{IJK}=g\varepsilon^{IJK}f_{abc}$
therefore gives a superconformal $\mathcal{N}=1$ theory at one-loop.
However, there are more general superpotentials satisfying the
one-loop finiteness conditions. To explore marginal deformations of
$\mathcal{N}=4$ SYM we consider the Leigh-Strassler superpotential
\cite{Leigh:1995ep} \be W=\frac{1}{3!}\lambda\varepsilon^{IJK}\tr
\left[\left[\Phi_{I},\Phi_{J}\right]\Phi_{K}\right]+ \frac{1}{3!}
h^{IJK}\tr \left[\left\{\Phi_{I},\Phi_{J}\right\}\Phi_{K}\right]\, ,
\label{deformed-superpotential}\ee where $h^{IJK}$ is totaly
symmetric. The coupling constants can now be written as
$C_{abc}^{IJK}=\lambda \varepsilon^{IJK}f_{abc}+h^{IJK}\tr
\left[\left\{T_{a},T_{b}\right\}T_{c}\right]$. The non-zero
 couplings are chosen to be  $h^{012}=\lambda (1-q)/(1+q)$ and
$h^{III}=2\lambda h/(1+q)$.
 In terms of the deformation parameters $q$ and $h$ the
superpotential (\ref{deformed-superpotential}) becomes
%%%%%
\be W=\frac{2\lambda}{1+q} \tr
\left[\Phi_{0}\Phi_{1}\Phi_{2}-q\Phi_{1}\Phi_{0}\Phi_{2}+
\frac{h}{3}\left(\Phi_{0}^{3}+\Phi_{1}^{3}+\Phi_{2}^{3}\right)\,
\right] .\label{qh-deformed} \ee
%%%%%
This  deformed superpotential will be our main focus.

The presence of $q$ and $h$ in the superpotential
(\ref{qh-deformed}) breaks the $SU(3)$ symmetry in the chiral
sector. What is left of the symmetry is a $Z_3\times Z_3$ symmetry.
The first $Z_3$ permutes the $\Phi$'s and the second takes
$\Phi_0\rightarrow \omega \Phi_0$, $\Phi_1\rightarrow \omega^2
\Phi_1$ and $\Phi_2\rightarrow  \Phi_2$, where $\omega$ is a third
root of unity.

The one-loop finiteness condition (\ref{finite}) is satisfied if
%%%5
\be\label{finite-condition}
g^2 = \frac{\lambda^{2}}{(1+q)^2}\left[(1+q)^2 + \left((1-q)^2 +2h^2
\right)\left(\frac{N^2 -4 }{N^2}\right)\right] .\label{couplings}
\ee
%%%%%
In the large-$N$ limit, which we consider, the relation
(\ref{couplings}) becomes even more simple. The one-loop finiteness
condition (\ref{finite}) also implies that the scalar field
self-energy contribution from the fermion loop is the same as in the
$\mathcal{N}=4$ scenario, due to the fact that the fermion loop has
the contraction $C_{acd}^{IKL}\bar{C}^{bcd}_{JKL}$. The parameters
in (\ref{finite-condition}) span a space within which there exists a
manifold, or perhaps just isolated points, $\beta (g,\lambda, q, h)
=0$ of superconformal theories to all loops \cite{Leigh:1995ep}. In
the limit $q\rightarrow 1$ and $h\rightarrow 0$ the $\mathcal{N}=4$
SYM is restored. Marginal deformations away from this fixed point
will be explored in the following sections by means of integrable
spin chains.

%%%%%%%%%%%%%%%%%%%%%%%%%%%%%%%%%%%%%%%%%%%%%%%%%%%%%%%%%%%%%%%%%%%%%%%%%%
%
%                      Section 3: Dilatation operator
%
%%%%%%%%%%%%%%%%%%%%%%%%%%%%%%%%%%%%%%%%%%%%%%%%%%%%%%%%%%%%%%%%%%%%%%%%%%

\section{Dilatation operator}
{}From the Leigh-Strassler deformation (\ref{qh-deformed}) of the
$\mathcal{N}=4$ SYM theory it is possible to obtain the dilatation
operator in the chiral sector. In this sector, the only contribution
is coming from the F-term in the Lagrangian, under the assumption
that the one-loop finiteness condition (\ref{finite}) is fulfilled.
The scalar field part of the F-term can be expressed in terms of the
superpotential as  \be \mathcal{L}_F = \biggr| \frac{\partial
W}{\partial \Phi_0}\biggr|^2 + \biggr| \frac{\partial W}{\partial
\Phi_1}\biggr|^2 + \biggr| \frac{\partial W}{\partial
\Phi_2}\biggr|^2. \ee
%%%%%
 Using $\phi_0$, $\phi_1$ and $\phi_2$ to denote
 the complex component fields, the Lagrangian becomes (omitting the overall factor $2\lambda /(1+q)$ in
(\ref{qh-deformed}))\bea \mathcal{L}_F &=& \tr\left[\phi_i
\phi_{i+1}\bar{\phi}_{i+1}\bar{\phi}_i
 -q\phi_{i+1}\phi_i\bar{\phi}_{i+1}\bar{\phi}_i
 - q^{*} \phi_i\phi_{i+1}\bar{\phi}_i\bar{\phi}_{i+1} \right]\nonumber \\
&+&\tr\left[ qq^{*}\phi_{i+1}\phi_i\bar{\phi}_i\bar{\phi}_{i+1}
-qh^{*} \phi_{i+1}\phi_i \bar{\phi}_{i+2}\bar{\phi}_{i+2}
-q^{*}h \phi_{i+2}\phi_{i+2}\bar{\phi_i}\bar{\phi}_{i+1}\right]\nonumber \\
&+&\tr\left[ h\phi_{i+2}\phi_{i+2}\bar{\phi}_{i+1}\bar{\phi}_i
+h^{*} \phi_i\phi_{i+1}\bar{\phi}_{i+2}\bar{\phi}_{i+2} +hh^{*}
\phi_{i}\phi_{i}\bar{\phi}_i\bar{\phi}_i \right], \eea where a
summation over $i=0,1,2$ is implicitly understood and the indices of
the
 fields $\phi_i$ are identified modulo three.
 In order to see how the dilatation operator acts on a general
operator $O=\psi^{i_1\ldots i_L}\tr \phi_{i_1}\ldots \phi_{i_L} $ to
first loop order in the planar limit we calculate the Feynman graphs
and regularize in accordance with
\cite{Roiban:2003dw,Berenstein:2004ys}. The vector space, spanning
these operators, can be mapped to the vector space of a spin-1 chain
(see \cite{Minahan:2002ve} for details). We define the basis states
$|0\rangle$, $|1\rangle$ and $|2\rangle$ for the spin chain which
correspond to the fields $\phi_0$, $\phi_1$ and $\phi_2$. By
introducing the operators $E_{ij}$, which act on the basis states as
$E_{ij}|k\rangle=\delta_{jk}|i\rangle$, the dilatation operator can
be written as a spin-chain Hamiltonian with nearest-neighbour
interactions, {\it i.e.} $\Delta=\sum_l H^{l,l+1}$ where \bea
\label{ekv:dil} H^{l,l+1} &=& E_{i,i}^{l} E_{i+1,i+1}^{l+1}  -q
E_{i+1,i}^{l} E_{i,i+1}^{l+1}  -q^{*}
E_{i,i+1}^{l} E_{i+1,i}^{l+1} \nonumber \\
&+&qq^{*} E_{i+1,i+1}^{l}   E_{i,i}^{l+1}  -qh^{*}
 E_{i+1,i+2}^{l} E_{i,i+2}^{l+1}   -q^{*}h
E_{i+2,i+1}^{l} E_{i+2,i}^{l+1}  \nonumber
\\&+&hE_{i+2,i}^{l} E_{i+2,i+1}^{l+1}
+h^{*}E_{i,i+2}^{l} E_{i+1,i+2}^{l+1} +hh^{*}E_{i,i}^{l}
E_{i,i}^{l+1}\, . \label{spin-chain-Hamiltonian} \eea The direct
product between the operators $E_{ij}$ is suppressed. If we use the
convention \be |0\rangle = \left(\begin{array}{c}
1 \\
0 \\
0
\end{array}\right)\qquad
|1\rangle = \left(\begin{array}{c}
0 \\
1 \\
0
\end{array}\right)\qquad
|2\rangle = \left(\begin{array}{c}
0 \\
0 \\
1
\end{array}\right),
\ee the Hamiltonian can be expressed as the matrix \be
H^{l,l+1}=\pmatrix{
  h^*h & 0 & 0 & 0 & 0 & h & 0 & -q^*h & 0 \cr
  0 & 1 & 0 & -q^* & 0 & 0 & 0 & 0 & h^* \cr
  0 & 0 & q^*q & 0 & -qh^* & 0 & -q & 0 & 0 \cr
  0 & -q & 0 & q^*q & 0 & 0 & 0 & 0 & -qh^* \cr
  0 & 0 & -q^*h & 0 & h^*h & 0 & h & 0 & 0 \cr
  h^* & 0 & 0 & 0 & 0 & 1 & 0 & -q^* & 0 \cr
  0 & 0 & -q^* & 0 & h^* & 0 & 1 & 0 & 0 \cr
  -qh^* & 0 & 0 & 0 & 0 & -q & 0 & q^*q & 0 \cr
  0 & h & 0 & -q^*h & 0 & 0 & 0 & 0 & h^*h } \, .
\ee

We will now search for special values of the parameters $h$ and $q$
for which the spin-chain Hamiltonian (\ref{spin-chain-Hamiltonian})
is integrable. When $h$ is absent, the analysis simplifies
considerably, because the usual S-matrix techniques can be used
\cite{Berenstein:2004ys,Freyhult:2005ws,Staudacher:2004tk}. The
existence of a homogeneous eigenstate, an eigenstate of the form
 $\mid \! a\rangle\otimes \mid \! a\rangle\ldots \otimes \mid \!
a\rangle$, is crucial for the S-matrix techniques to work. From this
reference state, excitations can be defined.  In this context, the
state $|a\rangle$ is a pure state, that is, one of the states
$|0\rangle$, $|1\rangle$ or $|2\rangle$.

When $h$ is non-zero, the analysis become significantly harder. The
only values for the parameters, for which it is possible to define a
homogeneous eigenstate are $q=1+e^{i2\pi n/3}h$
 or $q=-1$ and $h=e^{i2\pi n/3}$, where $n$ is an arbitrary integer.
 In these cases the homogeneous
 eigenstates are
\be
\mid \! a\rangle  = \mid \! 0\rangle +
 e^{\frac{i2\pi m}{3}}\mid \!
1\rangle +e^{-\frac{i2\pi m}{3}}\mid \! 2\rangle, \qquad m \in Z\, . \ee
Clearly, the two $Z_3$ symmetries are manifest. For $q=1+h e^{i2\pi
n/3}$, the eigenvalues are zero, thus the corresponding states are
protected. This case is related to the q-deformed Hamiltonian by a
simple change of variables. We introduce a new basis
 \bea
\label{base-shift} |0\rangle &=& \frac{e^{\frac{i2\pi
n}{3}}}{\sqrt{3}}(|\tilde{0}\rangle + |\tilde{1}\rangle
 + \tilde{2}\rangle)\, ,
   \nonumber \\
|1\rangle &=& \frac1{\sqrt{3}}(|\widetilde{0}\rangle +
e^{\frac{i2\pi}{3}}|\widetilde{1}\rangle
+ e^{-\frac{i2\pi}{3}}|\widetilde{2}\rangle)\, ,  \nonumber \\
|2\rangle &=& \frac1{\sqrt{3}}(|\widetilde{0}\rangle +
e^{-\frac{i2\pi}{3}}|\widetilde{1}\rangle +
e^{\frac{i2\pi}{3}}|\widetilde{2}\rangle), \eea
 where $n$ is an integer. It will shortly be shown that
 the phase shift in $|0\rangle$ will imply that a phase
  $e^{\pm i2\pi /3}$ can be transformed away from $h$.
 The Hamiltonian expressed in the new basis (\ref{base-shift}) takes the same form as
(\ref{ekv:dil}), but with new parameters $\tilde{q}$ and $\tilde{h}$
and an overall proportionality factor \be e^{\frac{i2\pi
}{3}}-qe^{-\frac{i2\pi }{3}}+he^{-\frac{i2\pi n}{3}}. \ee  The new
parameters $\tilde{q}$ and $\tilde{h}$ can then be expressed in
terms of the old parameters as \bea \label{ekv:transformation}
\tilde{q} &=&\frac{qe^{\frac{i2\pi }{3}}-e^{\frac{-i2\pi
}{3}}-he^{-\frac{i2\pi n}{3}}}
{e^{\frac{i2\pi }{3}}-qe^{\frac{-i2\pi }{3}}+he^{-\frac{i2\pi n}{3}}}\\
\tilde{h} &=&\frac{1-q+he^{-\frac{i2\pi n}{3}}}{e^{\frac{i2\pi
}{3}}-qe^{\frac{-i2\pi }{3}}+he^{-\frac{i2\pi n}{3}}}. \eea
%%%%%%%%%%%%%%%
 The case
$q=h e^{-i2\pi n/3}+1$  corresponds to the q-deformed case, and if
$h$'s imaginary part comes from the phase $e^{i2\pi n/3}$, the
remaining part is phase independent. This is in agreement with
reference~\cite{Berenstein:2000ux}. The integrable case $q$ equals a
phase will correspond to the case $q=he^{-i2\pi n/3}+1$ with $h=\rho
e^{i2\pi n/3}$ with $\rho$ and
 $q$  being real.
It is also clear that the case $q=-1$ and $h=e^{i2\pi n/3}$ is
related by the change of basis to a Hamiltonian of the form
\be
\label{hamiltonian-h0}
H=\sum_i 3\left[E_{22}^{i}E_{22}^{i+1} + E_{00}^{i}E_{00}^{i+1} +
E_{11}^{i}E_{11}^{i+1} \right]\, .
\ee
%%%
 This case looks perhaps
trivial, but it is not. The different eigenvalues equal $3 n$ with
$n=0,1,2, \ldots, L-2, L$. Note that the value $L-1$ is absent for
this periodic spin chain\footnote{Excitations can be created if
two states of the same number are next to each other. For example
the state $|112012\rangle$ has energy three and the next highest
energy state is $|111112\rangle$ with energy $4\times 3$. The state
with the highest energy, equals to $6\times 3$, is
$|111111\rangle$.}. The states have a large degree of degeneration.

For other values of $q$ and $h$, a reference state does not have a
precise meaning. Hence, we cannot adapt the S-matrix formalism.
Instead, we will try to find an R-matrix, from which the Hamiltonian
(\ref{spin-chain-Hamiltonian}) is obtainable. The existence of an
R-matrix $R(u)$, depending on the spectral parameter $u$, is
sufficient to ensure integrability. All R-matrices necessarily have
to satisfy the Yang-Baxter equation \be
R_{12}(u-v)R_{13}(u)R_{23}(v)=R_{23}(v)R_{13}(u)R_{12}(u-v).
\label{YangBaxter} \ee The Hamiltonian can be obtained from the
R-matrix through the following relation
%%%
\be \label{ekv:R} {\cal P}\frac{d}{du}R(u)|_{u=u_0}=H\,, \ee
%%%
where ${\cal P}$ is the permutation operator, with the additional
requirement $R(u_0)=\cal P$ for some point $u=u_0$.

%%%%%%%%%%%%%%%%%%%%%%%%%%%%%%%%%%%%%%%%%%%%%%%%%%%%%%%%%%%%%%%%%%%%%%%%%%
%
%                Section 4: A first look for integrability
%
%%%%%%%%%%%%%%%%%%%%%%%%%%%%%%%%%%%%%%%%%%%%%%%%%%%%%%%%%%%%%%%%%%%%%%%%%%

\section{A first look for integrability}
In this section, we will show how the transformation of basis
 (\ref{base-shift})
 combined with a position dependent phase shift, sometimes called a twist,
gives rise to new interesting cases of integrability. In
\cite{Berenstein:2004ys}, the $q$-deformed case was studied. It was
shown that for $q$ equals a root of unity, the phases can be
transformed away into the boundary conditions. Furthermore, it was
shown in \cite{Beisert:2005if} that the integrability properties do
not get affected for any $q=e^{i\beta}$, where $\beta$ is real. It
was also established that a generalised SYM Lagrangian deformed with
three phases $\gamma_i$, instead of just one variable, is
integrable. The deformed theory is referred to as the twisted (or
$\g$-deformed) SYM and the corresponding one-loop dilatation
operator in the three scalar sector is
\bea \label{ekv:lunin} H^{l,l+1} &=& \left[ E_{00}^{l}E_{11}^{l+1} +
E_{11}^{l}E_{22}^{l+1} + E_{22}^{l}E_{00}^{l+1} \right] \nonumber\\
&-&\left[ e^{i\g_1}E_{10}^{l}E_{01}^{l+1} +
e^{i\g_2}E_{21}^{l}E_{12}^{l+1} +
e^{i\g_3}E_{02}^{l}E_{20}^{l+1}\right]\nonumber\\
&-&\left[ e^{-i\g_1}E_{01}^{l}E_{10}^{l+1} +
e^{-i\g_2}E_{12}^{l}E_{21}^{l+1} + e^{-i\g_3}E_{20}^{l}E_{02}^{l+1}
\right]\nonumber\\ &+&\left[ E_{11}^{l}E_{00}^{l+1} +
E_{22}^{l}E_{11}^{l+1} + E_{00}^{l}E_{22}^{l+1}\right]\,. \eea

A natural question to ask is if the phases can also be transformed
away in a generic Hamiltonian of the form
(\ref{spin-chain-Hamiltonian}). If both $q$ and $h$ are present we
can not, at least in any simple way, transform away the phase of the
complex variables. However, when $q=r e^{\pm 2\pi i/3}$  it is
possible to do a position dependent coordinate transformation \be
\label{ekv:fas} |\tilde 0\rangle_k=e^{i2\pi/3}|0\rangle_k \,,\qquad
|\tilde 1\rangle_k=e^{i2k\pi/3}|1\rangle_k\,,\qquad \mbox{and}\qquad
|\tilde 2\rangle_k=e^{-i 2k\pi /3}|2\rangle_k \,,\ee as in
\cite{Berenstein:2004ys}\footnote{Note that the phase factor in
$|0\rangle$ is not position dependent, it was only added in order to
cancel the extra phase which would have appeared in front of the
terms having $h$ in them.} so that the phase of $q$ is transformed
away. Here, $k$ refers to the site of the spin-chain state. This
transformation changes the generators in the Hamiltonian as
%%%%%%%
\be \tilde{E}^{l}_{n,n+m}=e^{\frac{i2\pi ml}{3}}E^{l}_{n,n+m}\,. \ee
%%%%%%
This kind of transformation of basis generally results in twisted
boundary conditions. Thus, the periodic boundary condition
$|a\rangle_0=|a\rangle_L$ for the original basis becomes in the new
basis \be \label{twisted-boundary}
|\tilde{0}\rangle_0=|\tilde{0}\rangle_L \,,\qquad
|\tilde{1}\rangle_0=e^{\frac{i2\pi L}{3}}|\tilde{1}\rangle_L
\,,\qquad \mbox{and}\qquad |\tilde{2}\rangle_0=e^{\frac{-i2\pi
L}{3}}|\tilde{2}\rangle_L \,, \ee
%%%%%%%%%%
where $L$ is the length of the spin chain. A consequence is that the
system is invariant under a rotation of $q$ by introducing
appropriate twisted boundary conditions (\ref{twisted-boundary}). As
an example, the q-deformed Hamiltonian with periodic boundary
conditions with $q=h e^{i2\pi n/3}+1$ (see text above
(\ref{hamiltonian-h0})), is equivalent to $q e^{i2\pi m/3}=h
e^{i2\pi n/3}+1$ with twisted boundary conditions. Hence, the
following cases are integrable \be h=\rho e^{\frac{i2\pi n}{3}},
\;\;  \;\; q=(1+\rho)e^{\frac{i2\pi m}{3}}
 \hspace{0.7cm}\qquad\mbox{and} \qquad q=-e^{\frac{i2\pi m}{3}},\;\;\;
 h=e^{\frac{i2\pi
 n}3}\, ,
\ee where $\rho$ is  real and can take both negative and positive
values and $n$ and $m$ are arbitrary independent integers.

%%%%%%%%%%%%%%%%%%%%%%%%%%%%%%%

One can actually combine the twist transformation above with the
shift of basis (\ref{base-shift}) in a non-trivial way. This
combination will turn out to give a relation which maps the
Hamiltonian with arbitrary $q$ and vanishing $h$ into
 the Hamiltonian
with vanishing $q$ and arbitrary $h$. The periodic boundary
condition will, however, change for spin chains where the length is
not a multiple of three.

In terms of matrices the transformation can be represented as
follows. Let us represent the shift of basis (\ref{base-shift}) by
the matrix $T$ (with $n$ set to zero)
%%%%%%%%%%%%%%
\be T=\frac{1}{\sqrt{3}}\left(
\begin{array}{ccc}
  1 & 1 & 1   \\
  1 & e^{i2\pi/3} & e^{-i2\pi/3}   \\
  1 & e^{-i2\pi/3} & e^{i2\pi/3}   \\
  \end{array}\right) \,,
\ee
%%%%%%%%%%%%%
and the transformation matrix related to the phase shift
(\ref{ekv:fas}) by (but without the phase-shift in the zero state
$|0\rangle$) \be U_k=\left(
\begin{array}{ccc}
  1 & 0 & 0   \\
  0 & e^{i2\pi k/3} & 0 \\
  0 & 0 & e^{-i2\pi k/3}   \\
  \end{array}\right)\,.
\ee The transformation that takes the $q$-deformed to the
$h$-deformed Hamiltonian is then  \be \label{Htilde-transformation}
\widetilde{H}=T_1 H T_1^{-1} \,, \ee where \be T_1=(T\otimes
T)(U_{k}\otimes U_{k+1})(T^{-1}\otimes T^{-1})\,. \ee
 Acting with this transformation on the Hamiltonian
(\ref{ekv:dil}) we get the new Hamiltonian \bea \label{ekv:fil}
\widetilde{H}^{l,l+1} &=& q^*q E_{i,i}^{l} E_{i+1,i+1}^{l+1}
 -h q^*
 E_{i+1,i}^{l} E_{i,i+1}^{l+1}
-h^*q E_{i,i+1}^{l} E_{i+1,i}^{l+1} \nonumber \\
&+&hh^* E_{i+1,i+1}^{l}   E_{i,i}^{l+1} +h E_{i+1,i+2}^{l}
E_{i,i+2}^{l+1}
 +h^* E_{i+2,i+1}^{l} E_{i+2,i}^{l+1}  \nonumber
\\&-& q^* E_{i+2,i}^{l} E_{i+2,i+1}^{l+1}
-q E_{i,i+2}^{l} E_{i+1,i+2}^{l+1} +E_{i,i}^{l} E_{i,i}^{l+1}\,,
 \eea Up to an overall factor, the transformation
(\ref{Htilde-transformation}) change the couplings as
  \be \label{trans-qh} q\neq 0 \qquad \mbox{and} \qquad
h=0\qquad \Longleftrightarrow \qquad \tilde{q}=0 \qquad\mbox{and}
\qquad \tilde{h}=-1/q \ee In terms of states, the map
(\ref{Htilde-transformation}) generates the
 following change

\be \label{ekv:trans3} |a\rangle_{1+3k}\rightarrow
|a-1\rangle_{1+3k}\,, \;\;\;\; |a\rangle_{2+3k}\rightarrow
|a+1\rangle_{2+3k}\,, \;\;\;\; |a\rangle_{3k}\rightarrow
|a\rangle_{3k}\,, \ee where $a$ takes the values $0$, $1$ or $2$.
Let us investigate how  the transformation (\ref{ekv:fil})
affect the boundary conditions.
From equation (\ref{ekv:trans3}) we see that the original periodic
boundary conditions $|a\rangle_0=|a\rangle_L$ translate into
\be\label{ekv:transL1mod3}
|0^{new}\rangle_0=|2^{new}\rangle_L,\qquad
|1^{new}\rangle_0=|0^{new}\rangle_L \qquad \mbox{and} \qquad
|2^{new}\rangle_0=|1^{new}\rangle_L \, , \ee
 if the length $L$ of the spin chain is one modulo three and the opposite,
 $|0^{new}\rangle_0=|1^{new}\rangle_L$ {\it etc},
for the two modulo three case. If the length is a multiple of three the
boundary conditions remain the same.

If we start from the Hamiltonian of the $\g$-deformed SYM
(\ref{ekv:lunin}),  the transformation (\ref{Htilde-transformation})
leads to the Hamiltonian
 \bea \label{ekv:lunin2} H^{l,l+1} &=& \left[
E_{00}^{l}E_{11}^{l+1} + E_{11}^{i}E_{22}^{l+1} +
E_{22}^{i}E_{00}^{l+1} \right]  \nonumber\\
&-&\left[ e^{-i\g_{3-l}}E_{20}^{l}E_{21}^{l+1} +
e^{-i\g_{1-l}}E_{01}^{l}E_{02}^{l+1} +
e^{-i\g_{2-l}}E_{12}^{l}E_{10}^{l+1}\right]\nonumber\\
&-&\left[ e^{i\g_{3-l}}E_{02}^{l}E_{12}^{l+1} +
e^{i\g_{1-l}}E_{10}^{l}E_{20}^{l+1} +
e^{i\g_{2-l}}E_{21}^{l}E_{01}^{l+1} \right] \nonumber\\
&+&\left[ E_{00}^{l}E_{00}^{l+1} + E_{11}^{l}E_{11}^{l+1} +
E_{22}^{l}E_{22}^{l+1}\right] \,.
\eea
%%%%
This Hamiltonian describes
interactions which differ from systems we have previously
encountered, since here the interactions are site dependent. This
behavior shows up naturally in a non-commutative theory. In
\cite{Beisert:2005if}, it was discussed that the $\gamma$-deformed
SYM  corresponds to a form of  non-commutative deformation
of $\mathcal{N}=4$ SYM.

If all the $\g_i$ are equal, the Hamiltonian above will corresponds
to our original Hamiltonian (\ref{ekv:dil}) with $q=0$ and
$h=e^{i\theta}$. The associated R-matrix is \bea \label{Rmatrix1}
 R(u) &=&\left[ E_{01}^{i}E_{10}^{i+1} + E_{12}^{i}E_{21}^{i+1}
+
E_{20}^{i}E_{02}^{i+1} \right]  \nonumber\\
&-&u e^{-i\theta}\left[ E_{20}^{i}E_{21}^{i+1} +
E_{01}^{i}E_{02}^{i+1} +
E_{12}^{i}E_{10}^{i+1}\right]\nonumber\\
&-&u e^{i\theta}\left[ E_{12}^{i}E_{02}^{i+1} +
E_{20}^{i}E_{10}^{i+1} +
E_{01}^{i}E_{21}^{i+1} \right] \nonumber\\
&+&\left[ E_{00}^{i}E_{00}^{i+1} + E_{11}^{i}E_{11}^{i+1} +
E_{22}^{i}E_{22}^{i+1}\right]\nonumber\\
&+&(1-u)\left[ E_{10}^{i}E_{01}^{i+1} + E_{21}^{i}E_{12}^{i+1} +
E_{02}^{i}E_{20}^{i+1} \right]\,.  \eea We have checked explicit
that (\ref{Rmatrix1}) satisfies the Yang-Baxter equation. This means
that the theory is integrable! \\ \\
 In the rest of this section we will discuss the spectrum when the
spin-chain Hamiltonian (\ref{ekv:dil}) is either $q$-deformed or
$h$-deformed. Figure \ref{fig:spektra1} shows the spectrum for a
three-site spin-chain Hamiltonian. The left graph shows how the
energy depends on the phase $\phi$, with $q=e^{i\phi}$ and $h=0$.
The right graph shows instead how the eigenvalues vary as a function
$\tilde{\theta}$, when $\tilde{h}=e^{i\tilde{\theta}}$ and
$\tilde{q}=0$.
%%%%%%%%%%%%%%%
\FIGURE[t]{\label{fig:spektra1}
\parbox{7cm}{\centering\includegraphics[height=6cm]{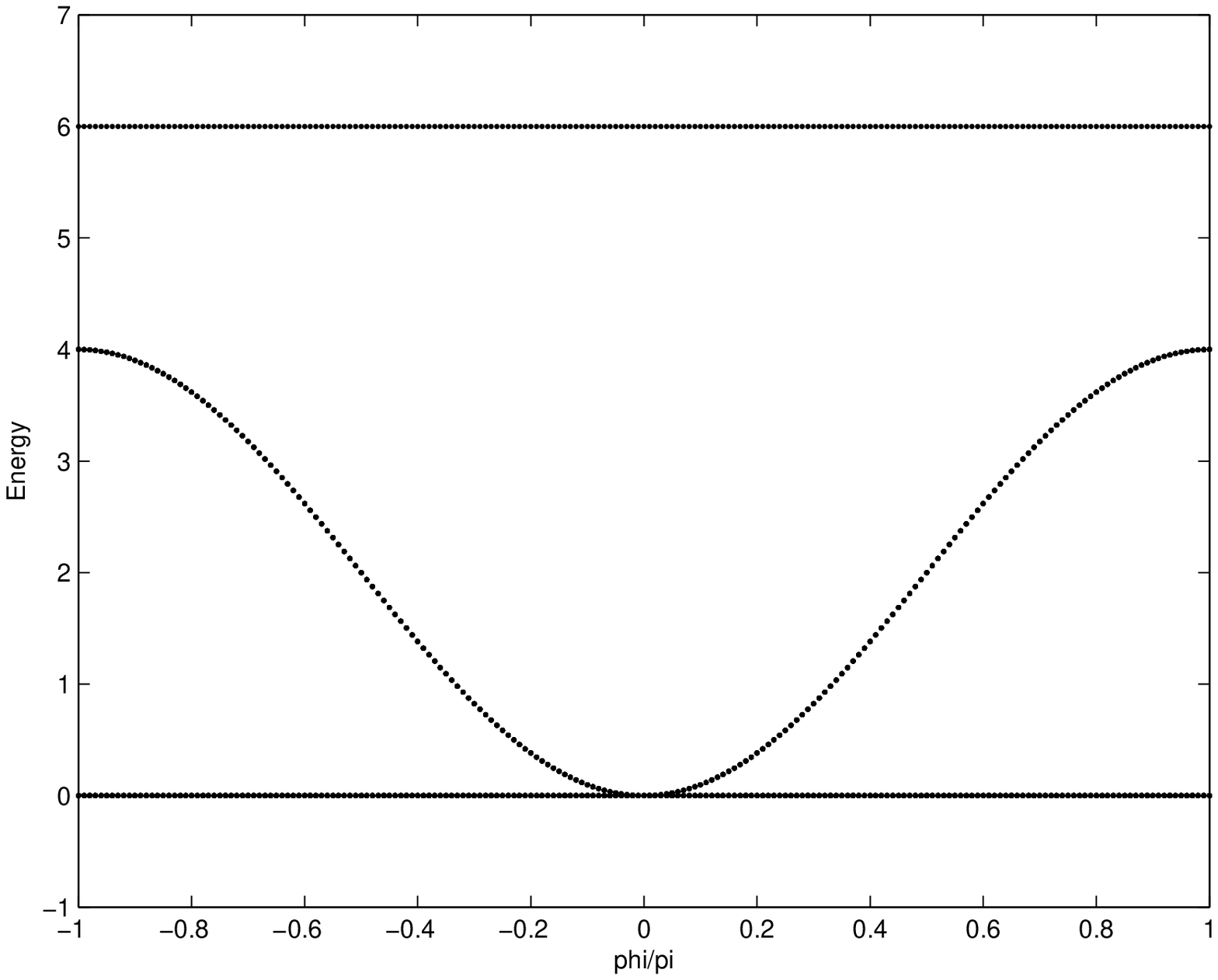}}
$\;\;$
\parbox{7cm}{\centering\includegraphics[height=6cm]{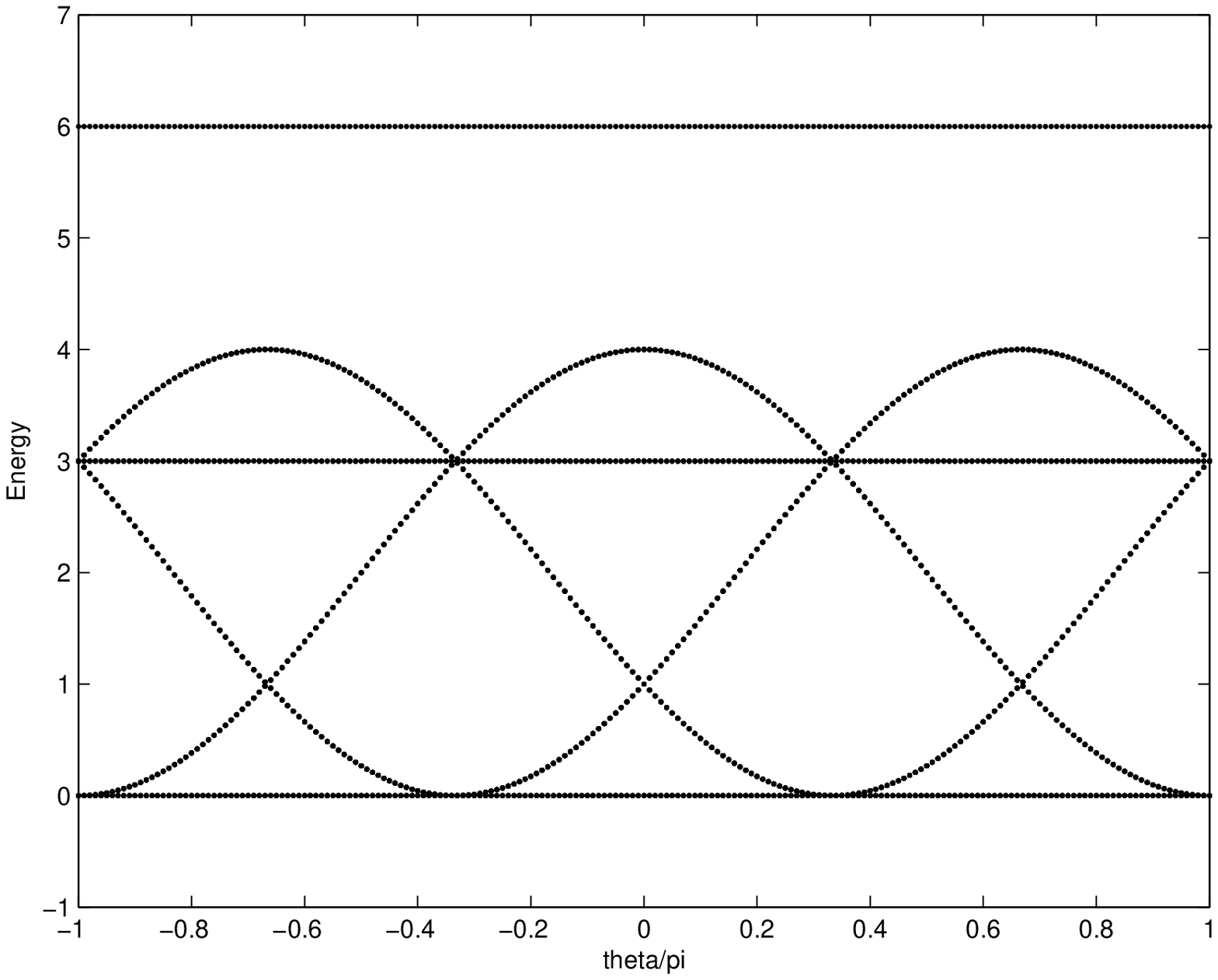}}
\caption{Spin chain with three sites. The left graph shows the
energy spectrum as a function of the phase $\phi$, when
 $q=e^{i\pi \phi}$ and $h=0$. The right graph shows the
spectrum as a function of the phase $\tilde{\theta}$, when
$\tilde{h}=e^{i\tilde{\theta}}$ and $\tilde{q}=0$. } }
%%%%%%%%%
Figure \ref{fig:spektra} shows the same spectra for a four-site spin
chain.
%%%%%%%%%
\FIGURE[t]{\label{fig:spektra}
\parbox{7cm}{\centering\includegraphics[height=6cm]{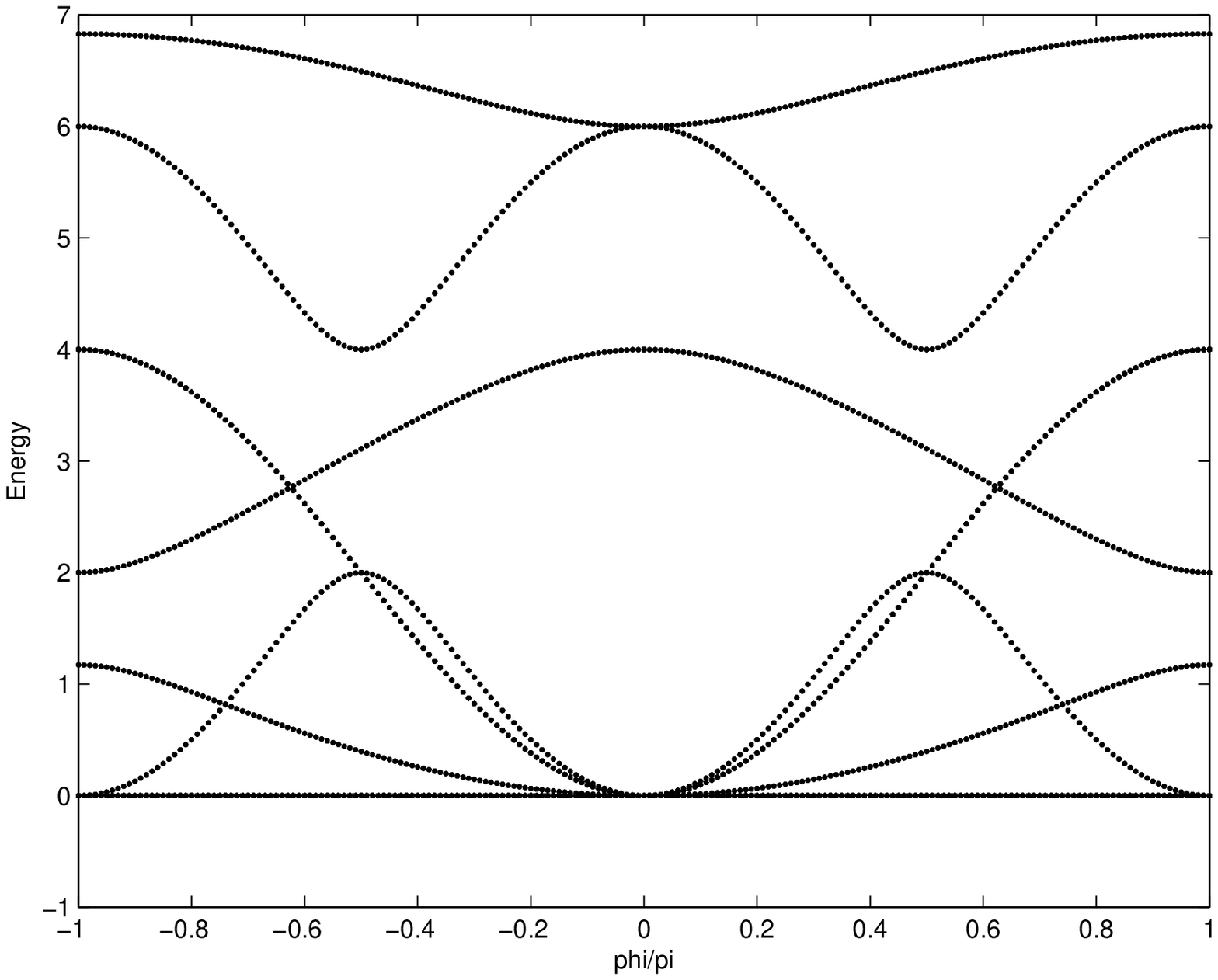}}
$\;\;$
\parbox{7cm}{\centering\includegraphics[height=6cm]{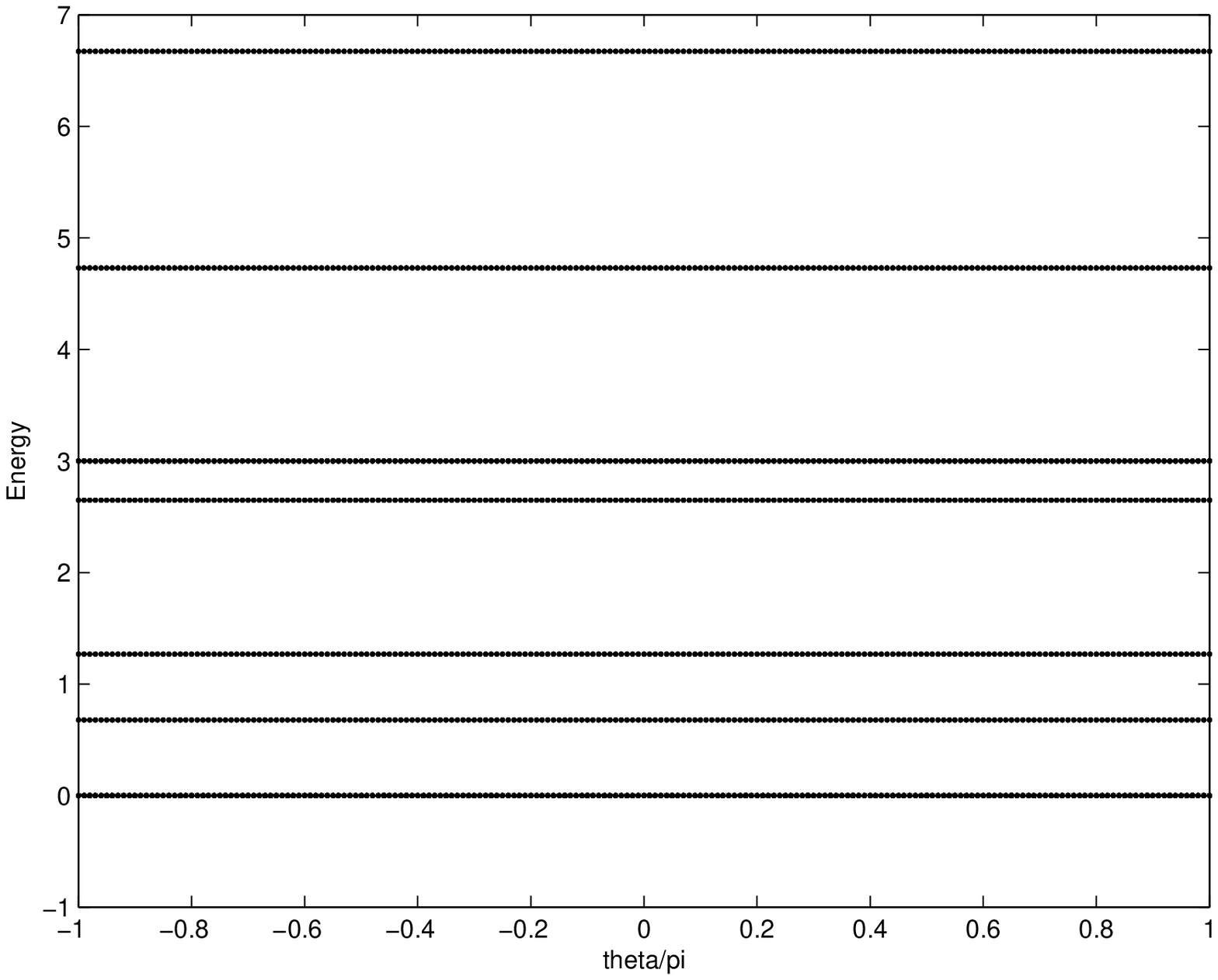}}
\caption{Spin chain with four sites. The left graph shows the energy
spectrum as a function of the phase $\phi$, when
 $q=e^{i\pi \phi}$ and $h=0$. The right graph shows the
spectrum as a function of the phase $\tilde{\theta}$, when
$\tilde{h}=e^{i\tilde{\theta}}$ and $\tilde{q}=0$. }}
%%%%%%%%%%%
All graphs contain energies which are the eigenvalues of several
states. Highly degenerate states are generally a sign of
integrability because they reflect a large number of symmetries in
the theory.

Let us start by explaining the spectra in Figure \ref{fig:spektra1}.
When $h$ is zero there is only one sinus curve while when $q$ is
zero there are three sinus curves. The reason is the transformation
(\ref{trans-qh}), since it maps $q=e^{i\phi}$ and $h=0$ into
$\tilde{h}=e^{i\tilde{\theta}}$ and $\tilde{q}=0$ with the relation
of the phases $\tilde{\theta}=\pi - \phi + 2\pi n/3$. Therefore, for
each value of $q$ there exist several values of $\tilde{h}$ which
differ by a phase $2\pi/3$. For $q=0$, there is a state, independent
of the phase, with energy three. This state is absent for $h=0$. One
example of such a state is $|000\rangle-|111\rangle$. The ``inverse"
transformation, see (\ref{ekv:trans3}), of this state is
$|120\rangle-|201\rangle$, which is zero due to periodicity.

The four-site spin chain (see Figure \ref{fig:spektra}) differs
substantially from the spin chain with three sites. The case $q=0$
is completely phase-independent. The reason is  the
boundary conditions. Actually, spin chains with the number of sites
differing from multiples of three will have spectra which do not
depend on the phase. It will just coincide with the spectra for the
case $q=e^{-i2\pi/3}$ and $h=0$. Starting with the case $q$ equal to
a root of unity it is possible to make a transformation, changing
the boundary conditions, such that the phase of $q$ is removed
\cite{Berenstein:2004ys}. The change in the boundary conditions is
then
%%%%
\be \label{bc-twist} |0^o\rangle_0=|0^o\rangle_L \,, \qquad
|1^o\rangle_0=e^{i\Phi}|1^o\rangle_L \qquad \mbox{and}\qquad
|2^o\rangle_0=e^{-i\Phi}|2^o\rangle_L \,, \ee where $\Phi$ is a
phase factor, the exact form of which is not important for our
purposes. The effect (\ref{bc-twist}) has on the boundary conditions
(\ref{ekv:transL1mod3}) is, when $L$ is one modulo three, \be
|0^{new}\rangle_0 = |2^{new}\rangle_L\,,\qquad |1^{new}\rangle_0 =
e^{i\Phi}|0^{new}\rangle_L \qquad \mbox{and}\qquad
|2^{new}\rangle_0=e^{-i\Phi} |1^{new}\rangle_L \,. \ee

 If we make the shift $|1^{new}\rangle \rightarrow
e^{i\Phi}|1^{new}\rangle$ we see that this corresponds to the
boundary conditions (\ref{ekv:transL1mod3}).
 The same procedure can be made when $L$ is two modulo three. This means that any $q$ equal
  to root of unity\footnote{$q=e^{i\phi}$ is a root of unity iff
  $n\phi = 0\, \mbox{mod}\, 2\pi$ for $n$ an integer. The phase is then $\phi=2\pi p/n$ where $p$ is an integer. }
  can be mapped to any
$\tilde{h}$ with the phase $\tilde{\theta}=\pi + 2\pi p/n+2\pi m/3$.
All values of $h$ will then give the same energy spectrum due to the
fact that
 $p$,$n$ and $m$ are arbitrary integer numbers, so the possible
 values of $\tilde{\theta}$ will in principle fill up the real axis. This implies that the
energy must be the same for all values of $\tilde{\theta}$. For
$q=e^{-i2\pi/3}$ and $h=0$ there is a direct map (see
(\ref{ekv:transformation})) to the case $q=0$ and $h=-e^{2\pi m/3}$
 which does not change the boundary conditions.
The energy spectra for these two cases must be the same.
Consequently, the spectra for ``all'' points coincide with the
spectrum of $q=e^{-i2\pi/3}$ and $h=0$. The fact that the shape of
the eigenvalue distribution changes drastically depending on how
many sites there are suggests that a well-defined large L-limit does
not exist. However, it might still be possible to find a
well-defined large L-limit if only  L-multiples of three is
considered.

\section{R-matrix}
 We will now try to make a general ansatz for an R-matrix which has the
possibility to give rise to our Hamiltonian (\ref{ekv:dil}). A
linear ansatz will turn out to lead to the cases we found in the previous
section. To
find a new solution the ansatz need to be more complicated, for
instance consisting of hyper-elliptic functions. We are interested
in an  R-Matrix of the following form \bea R(u) &=& aE_{i,i}\otimes
E_{i,i}+bE_{i,i}\otimes E_{i+1,i+1}+
\bar{b}E_{i+1,i+1} \otimes  E_{i,i} \nonumber \\
&+&cE_{i,i+1}\otimes E_{i+1,i} +\bar{c}E_{i+1,i}\otimes E_{i,i+1}
\nonumber \\
&+&dE_{i+1,i+2}\otimes E_{i,i+2} + \bar{d}E_{i+2,i+1}\otimes
E_{i+2,i}
\nonumber \\
&+&eE_{i+2,i}\otimes E_{i+2,i+1}+\bar{e}E_{i,i+2}\otimes
E_{i+1,i+2},
\eea
where the coefficients are functions of a
spectral parameter $u$.

Written  on matrix form the R-matrix is
\be\label{ekv:rmatrix}
R=\pmatrix{
  a & 0 & 0  & 0 & 0 & e & 0 & \bar{d} & 0 \cr
  0 & b & 0   & c & 0 & 0 & 0 & 0 & e \cr
  0 & 0 & \bar{b}  & 0 & d & 0 & \bar{c} & 0 & 0 \cr
  0 & \bar{c} & 0   & \bar{b} & 0 & 0 & 0 & 0 & d \cr
  0 & 0 & \bar{d}   & 0 & a & 0 & e & 0 & 0 \cr
  \bar{e} & 0 & 0   & 0 & 0 & b & 0 & c & 0 \cr
  0 & 0 & c &   0 & \bar{e} & 0 & b & 0 & 0 \cr
  d & 0 & 0 &   0 & 0 & \bar{c} & 0 & \bar{b} & 0 \cr
  0 & e & 0  & \bar{d} & 0 & 0 & 0 & 0 & a }\, .
\ee
A natural first step to look for a R-matrix solution is
to make a linear ansatz which will give the Hamiltonian (\ref{ekv:dil})
 as in (\ref{ekv:R}).

The Hamiltonian can also be defined through the permuted
$\mathfrak{R}$-matrix \be \mathfrak{R}\equiv {\cal P}R, \ee where
$\cal P$ is the $9\times9$ permutation matrix.

If  $R(u)|_{u=u_0}={\cal P}$
or $\mathfrak{R}(u)|_{u=u_0}={\cal P}$,
 the Hamiltonian is obtained as
\be \label{ekv:hu}
H={\cal P}\frac{d}{du}R(u)|_{u=u_0}
\hspace{0.4cm} \mbox{or}\hspace{0.4cm}
 H={\cal P}\frac{d}{du}\mathfrak{R}(u)|_{u=u_0}.
\ee
The linear
ansatz below has the property that it gives the Hamiltonian (\ref{ekv:dil})
in accordance with the first formula in (\ref{ekv:hu})
\bea \label{ekv:linear}
a(u)&=&(h^*h-k)u +\alpha\, , \qquad b(u)=-qu \,,
\qquad \bar{d}(u)=-q^*h u\,, \nonumber  \\
c(u)&=&(q^*q-k)u +\alpha\,, \hspace{0.8cm}\bar{b}(u)=-q^* u\,,
\hspace{0.7cm}e(u)=h u \,,  \nonumber \\
\bar{c}(u)&=&(1-k)u +\alpha\,,\hspace{1.1cm}
 d(u)=h^*u \,,\hspace{1cm}\bar{e}(u)=-qh^*u \,,
\eea
with $k$ and $\alpha$ being free parameters, the Yang-Baxter equations
turn out to be independent of $\alpha$ while they demand $k$ to be
$k=\frac1{2}\left(1+h^*h+q^*q\right)$.
Inserting the linear ansatz in the
Yang-Baxter equation we find that the equation is satisfied
either if
\be
q=e^{i\phi}\; \mbox{and}\; h=0 \;\;\; \mbox{or} \;\;\;
q=0 \; \mbox{and} \; h= e^{i\theta}\; ,
\ee
where  $\phi$ and $\theta$ can be any phase,
 or if the following equations holds
\bea
e^{i3\phi}r&=&\left(1 +\rho e^{i3\theta} \right)\,, \\
e^{-i3\phi}r&=&\left(1 +\rho e^{-i3\theta} \right)\,, \\
r&=& \pm(1\pm \rho) \, ,
\eea
where we used the notation that $q=re^{i\phi}$ and
$h=\rho e^{i\theta}$ and  let $r$ and $\rho$
be any real numbers.
Here we immediately see that the relations between the
real parts of $q$ and $h$ are given by the last equation, hence we only
need to consider which angles are not in
contradiction to that.
The solution is
\be
q=r e^{i2\pi n/3}\,,\hspace{1cm} h=(1+r)e^{i2\pi m/3}\, ,
\ee
where we once again let $r$ take any real number.
Now we would like to see whether there exist solutions if
an ansatz is made with the permuted version of the R-matrix ansatz (\ref{ekv:linear}).
We obtain
\bea
&&a(u)=(h^*h-k)u +\alpha \,, \hspace{1cm} c(\mu)=-q^* u\,,
\hspace{1cm}e(u)=h u\,, \nonumber  \\
&&b(u)=(1-k)u +\alpha \,,\hspace{1.4cm}\bar{c}(u)=-q u\,,
\hspace{1.2cm}\bar{d}(u)=-q^*h u\,,   \nonumber \\
&&\bar{b}(u)=(q^*q-k)u +\alpha\,,\hspace{1cm}
 d(u)=-qh^* u \,,\hspace{1cm}\bar{e}(u)=h^* u \, .
\eea The conditions from the Yang-Baxter equation read \be
q^*=-q^2\,, \hspace{1cm} h^*=h^2 \, , \ee with no restriction on $k$
and $\alpha$. The only solution to this is \be q=-e^{2\pi n/3}\,,
\hspace{1cm} h=e^{2\pi m/3} \, , \ee (or $q=0$ and $h=0$). This is
the other type of solution we expected from the last section. The
one corresponding to $q=-1$ and $h=e^{i2\pi m/3}$ and that one but
with twisted boundary conditions. Hence a R-matrix with a linear
dependence on the spectral parameter $u$ can not give us more
integrable cases than already found. We need a more general R-matrix
solution to find new interesting cases.
\subsection{Symmetries revealed}
In order to address the problem of finding the most general solution
for the R-matrix (\ref{ekv:rmatrix}) it is an advantage to make use of
the symmetries. We choose the representation
%%%
\be \label{ekv:Rmatrix2}
R=\sum_{i=1}^3
(\o_i T_i\otimes S_i+\bar{\o}_i S_i\otimes T_i+\gamma_{i} E_i\otimes
E_{2i})\,.
\ee
%%%%
 All indices in this section are modulo three if not
otherwise stated. The generators $S_i$, $T_i$ and $E_i$ are \bea
S_1= \pmatrix{
  0 & 0 & 1   \cr
  e^{-\frac{i2\pi}{3}} & 0 & 0\cr
  0 &  e^{\frac{i2\pi}{3}}& 0
  },\!\! \qquad\!\!
S_2 = \pmatrix{
  0 & 0 & 1   \cr
  e^{\frac{i2\pi}{3}} & 0 & 0\cr
  0 &  e^{-\frac{i2\pi}{3}}& 0
  },\!\! \qquad \!\!
S_3 = \pmatrix{
  0 & 0 & 1 \cr
  1 & 0 & 0 \cr
  0 & 1 & 0
  },
\nonumber \\
T_1= \pmatrix{
  0 & e^{\frac{i2\pi}{3}} & 0   \cr
  0 & 0 &  e^{-\frac{i2\pi}{3}}\cr
  1 & 0 & 0}, \!\! \qquad \!\!
T_2 = \pmatrix{
  0 & e^{-\frac{i2\pi}{3}} & 0   \cr
  0 & 0 &  e^{\frac{i2\pi}{3}} \cr
  1 & 0 & 0
  },\!\! \qquad \!\!
T_3 = \pmatrix{
  0 & 1 & 0   \cr
  0 & 0 &  1 \cr
  1 & 0 & 0  },
\nonumber \\
E_1 = \pmatrix{
  1 & 0 & 0   \cr
  0 &e^{-\frac{i2\pi}{3}} & 0\cr
  0 & 0 &  e^{\frac{i2\pi}{3}}
  },\!\!\qquad \!\!
E_2 = \pmatrix{
  1 & 0 & 0   \cr
  0 &e^{\frac{i2\pi}{3}} & 0\cr
  0 & 0 &  e^{-\frac{i2\pi}{3}}
  },\!\!\qquad \!\!
E_3 = \pmatrix{
  1 & 0 & 0   \cr
  0 & 1 & 0\cr
  0 & 0 &  1
  }.
\eea How the functions in the R-Matrix (\ref{ekv:rmatrix}) are expressed
in terms of the functions $\o_i$, $\bar{\o}_i$ and $\gamma_{i}$ can be
found in Appendix (\ref{parameter}).
The generators are related by \be
\begin{array}{lll}
  S_kS_{l} = e^{-i\frac{2\pi (l-k)}{3}}T_{2k-l}\qquad & S_kT_l = E_{k-l}\qquad
  & S_kE_l =  e^{i\frac{2\pi l}{3}}S_{k+l} \\
  T_kS_l = e^{-\frac{i2\pi(l-k)}{3}}E_{l-k}\qquad  & T_kT_l
= e^{-i\frac{2\pi (l-k)}{3}}S_{2k-l} \qquad & T_kE_l = T_{k-l} \\
  E_kS_l = S_{k+l}\qquad  & E_kT_l = e^{i\frac{2\pi}{3}k}T_{l-k}\qquad  & E_kE_l = E_{k+l} \\
\end{array}\ee
Using these relations it is straightforward to obtain the
Yang-Baxter equations which can be found in Appendix (\ref{ekv:YB}).
A nice feature of these equations is that all of them,
 except the fourth, the fifth and the sixth, can be generated
 from the first equation through the cyclic permutations
$\o_{n+1}\rightarrow \bar{\o}_{n+1}\rightarrow \g_3$ and
$\g_2\rightarrow \o_n\rightarrow \bar{\o}_n\rightarrow \g_1
\rightarrow \o_{n+2}\rightarrow \bar{\o}_{n+2}$. The remaining three
equations are related to each other by the same cyclic permutation.
The structure of the equations (\ref{ekv:YB}) is similar to the
Yang-Baxter equations in the eight vertex model
\cite{Baxter:1978xr,Gomez:1996az}
%%%
\be \label{ekv:8vert} \o_n \o_l'
\o_j''-\o_l\o_n'\o_k''+\o_j\o_k'\o_n''-\o_k\o_j'\o_l''=0\;,
\ee
%%%
 for
all cyclic permutations ($j,k,l,n$) of (1,2,3,4). These equations
can neatly be represented by writing the elements in rectangular
objects \be
\begin{array}{|c|c|c|c|}
\hline {\bf{\o_n}} & \o_l & {\bf\o_j} & \o_k \\
\hline \o_l & {\bf\o_n} & \o_k &  {\bf\o_j} \\
\hline {\bf\o_j} & \o_k & {\bf\o_n} & \o_l \\
\hline \o_k & {\bf\o_j} & \o_l & {\bf\o_n} \\
\hline  {\bf\o_n} & \o_l & {\bf\o_j} & \o_k \\
\hline
\end{array}\,.
\ee
Note the beautiful toroidal pattern.
The object above should be interpreted as follows. The first three
 rows represent the equation (\ref{ekv:8vert}) with the first column
representing the first term in (\ref{ekv:8vert})
\be
\begin{array}{|c|c|c|c|}
\hline {\bf\o_n} \\
\hline \o_l  \\
\hline {\bf\o_j} \\
\hline
\end{array} =\o_n \o_l' \o_j''\,,
\ee and the next column is equal to the second term in
(\ref{ekv:8vert}) \be
\begin{array}{|c|c|c|c|}
\hline \o_l \\
\hline {\bf\o_n}  \\
\hline \o_k \\
\hline
\end{array} =-\o_l \o_n' \o_k''\,.
\ee The next three rows represent another equation  of eight vertex
model  \be
\begin{array}{|c|c|c|c|}
\hline \o_l & {\bf\o_n} & \o_k &  {\bf\o_j} \\
\hline {\bf\o_j} & \o_k & {\bf\o_n} & \o_l \\
\hline \o_k & {\bf\o_j} & \o_l & {\bf\o_n} \\
\hline
\end{array}=\o_l \o_j' \o_k''-\o_n\o_k'\o_j''+
\o_k\o_n'\o_l''-\o_j\o_l'\o_n''=0\,. \ee Our equations can also be
represented in terms of similar rectangular objects, with the same
toroidal pattern \be
\begin{array}{|c|c|c|c|c|c|}
\hline  {\bf \o_2} & \o_1 & {\bf\bar{\o}_2} & \g_1 & {\bf\g_3} &
 \bar{\o}_3 \\
\hline \o_1 & {\bf\o_2} & \g_1 & {\bf\bar{\o}_2} & \bar{\o}_3 & {\bf\g_3} \\
\hline {\bf\g_3} & \g_1 & \bf{\o_2} & \bar{\o}_3 & {\bf\bar{\o}_2} & \o_1 \\
\hline \g_1 & {\bf\g_3} & \bar{\o}_3 & {\bf\o_2} & \o_1 & {\bf\bar{\o}_2} \\
\hline {\bf\bar{\o}_2} & \bar{\o}_3 & {\bf\g_3} & \o_1 & \bf{\o_2} & \g_1 \\
\hline
\end{array}\,.
\ee The first three rows give the second equation in (\ref{ekv:YB})
with $n=3$. The next three rows are the seventh equation in
(\ref{ekv:YB}) with $n=1$. This suggests that the system of
equations (\ref{ekv:YB}) should have a nice solution, just like  the
eight vertex model. The first row determines the rest of the
entries, thus all equations can be represented with just the upper
row. Hence, all the 36 equations can be represented by the following
rows \bea &&\begin{array}{|c|c|c|c|c|c|} \hline \o_{n+1} &
\,\;\o_n\;\, & \bar{\o}_{n+1} & \,\;\g_1\;\, &\, \;\g_3\;\, &
 \bar{\o}_{n+2} \\
\hline
\end{array} \qquad
\begin{array}{|c|c|c|c|c|c|}
\hline \o_{n+1} & \;\o_n\;\, & \;\bar{\o}_n\;\, & \;\g_2\;\, &\; \g_1\;\, &
\bar{\o}_{n+1} \\
\hline
\end{array} \nonumber
\\
&&\begin{array}{|c|c|c|c|c|c|}
\hline \,\;\o_n \;\, & \o_{n+1} &\,\; \bar{\o}_n\;\, &\, \;\g_2\;\,
 & \,\;\g_3\;\, &
\bar{\o}_{n+2} \\
\hline
\end{array} \qquad
\begin{array}{|c|c|c|c|c|c|}
\hline \,\;\o_2\;\, &\, \;\g_2\;\, &\, \;\o_1\;\, & \,\;\g_1\;\,
&\,\; \o_3\;\, &
\,\;\g_3\;\, \\
\hline
\end{array}  \\
&&\begin{array}{|c|c|c|c|c|c|}
\hline \;\;\bar{\o}_2 \;\,& \,\;\o_2\;\, &\, \;\bar{\o}_1\;\, &\, \;\o_1\;\,
&\;\; \bar{\o}_3\;\, &\; \;\o_3\;\, \\
\hline
\end{array}\qquad
\begin{array}{|c|c|c|c|c|c|}
\hline \,\;\g_1\;\, & \,\;\bar{\o}_2\;\, &\, \;\g_2\;\, & \,\;\bar{\o}_1\;\,
 &\,\; \g_3\;\, &
\,\;\bar{\o}_{3}\;\, \\
\hline
\end{array} \nonumber \,.
\eea The solution to the eight vertex model is a product of theta
functions. The cyclicity and periodicity properties of the eight
vertex model is mirrored into the rectangular object. Due to the
combination of addition theorems for theta functions and the
intrinsic properties of the equations, the rectangular objects make
it easy to see if an ansatz solves all the equations. We believe
that the addition theorems for theta functions generating the
solution of the eight vertex model should be possible to generalize
to any even sized rectangular object. It would then be interesting
to see if those equations are related to an R-matrix of arbitrary
dimension.
%%%%%%%%%%%%%%%%%%%%%%%%%%%%%%%%%%%5
%%%%%%%%%%%%%%%%%%%%%%%%%%%%%%%%%%%
\subsection{A hyperbolic solution}
If the following ansatz $\o_i=e^{uQ_i}$, $\bar{\o}_i=e^{u\bar{Q}_i}$
and $\g_i=e^{uK_i}$, where we let $Q_i$, $\bar{Q}_i$ and $K_i$ be arbitrary
constants is made, it leads us to the following solution
\bea
&\o_1=e^{Q_1 u}\,, \;\;\;\;\; \bar{\o}_1 =e^{Q_2 u}\,,\;\;\;\;\;\g_1=e^{Q_2 u}
\,,\nonumber \\
&\o_2=e^{Q_2 u}\,, \;\;\;\;\; \bar{\o}_2=e^{Q_1 u}\,,\;\;\;\;\; \g_2=e^{Q_1 u}
\,, \\
&\o_3=e^{Q_3 u}\,, \;\;\;\;\; \bar{\o}_3=e^{Q_3 u}\,,
\;\;\;\;\; \g_3=e^{Q_3 u}\,.
\nonumber\,,
\eea
The following Hamiltonian is obtained from the above R-matrix solution
\bea \label{ekv:dil2} H^{l,l+1} &=& E_{i,i}^{l}\otimes
E_{i+1,i+1}^{l+1}  +e^{i\phi} E_{i+1,i}^{l}\otimes E_{i,i+1}^{l+1}
 +e^{-i\phi} E_{i,i+1}^{l}\otimes E_{i+1,i}^{l+1} \nonumber \\
&+& E_{i+1,i+1}^{l} \otimes  E_{i,i}^{l+1}  +e^{-i\phi}
 E_{i+1,i+2}^{l}\otimes E_{i,i+2}^{l+1}   +e^{i\phi}
E_{i+2,i+1}^{l}\otimes E_{i+2,i}^{l+1}  \nonumber
\\&+& e^{-i\phi}E_{i+2,i}^{l}\otimes E_{i+2,i+1}^{l+1}
+e^{i\phi}E_{i,i+2}^{l}\otimes E_{i+1,i+2}^{l+1}
+E_{i,i}^{l}\otimes E_{i,i}^{l+1}.
 \eea
where $e^{i\phi}=(Q_2e^{i2\pi/3}+Q_1e^{-i2\pi/3})/(Q_1^2+Q_2^2-Q_1Q_2)$
(we put $Q_3$ to zero because it does not give us any more information).
Here we also made use of the fact that the Hamiltonian obtained from the
procedure (\ref{ekv:R}) can be rescaled plus that something proportional to
the identity matrix can be added.
\FIGURE[t]{\label{fig:spektra70}
\parbox{7cm}{\centering\includegraphics[height=5.5cm]{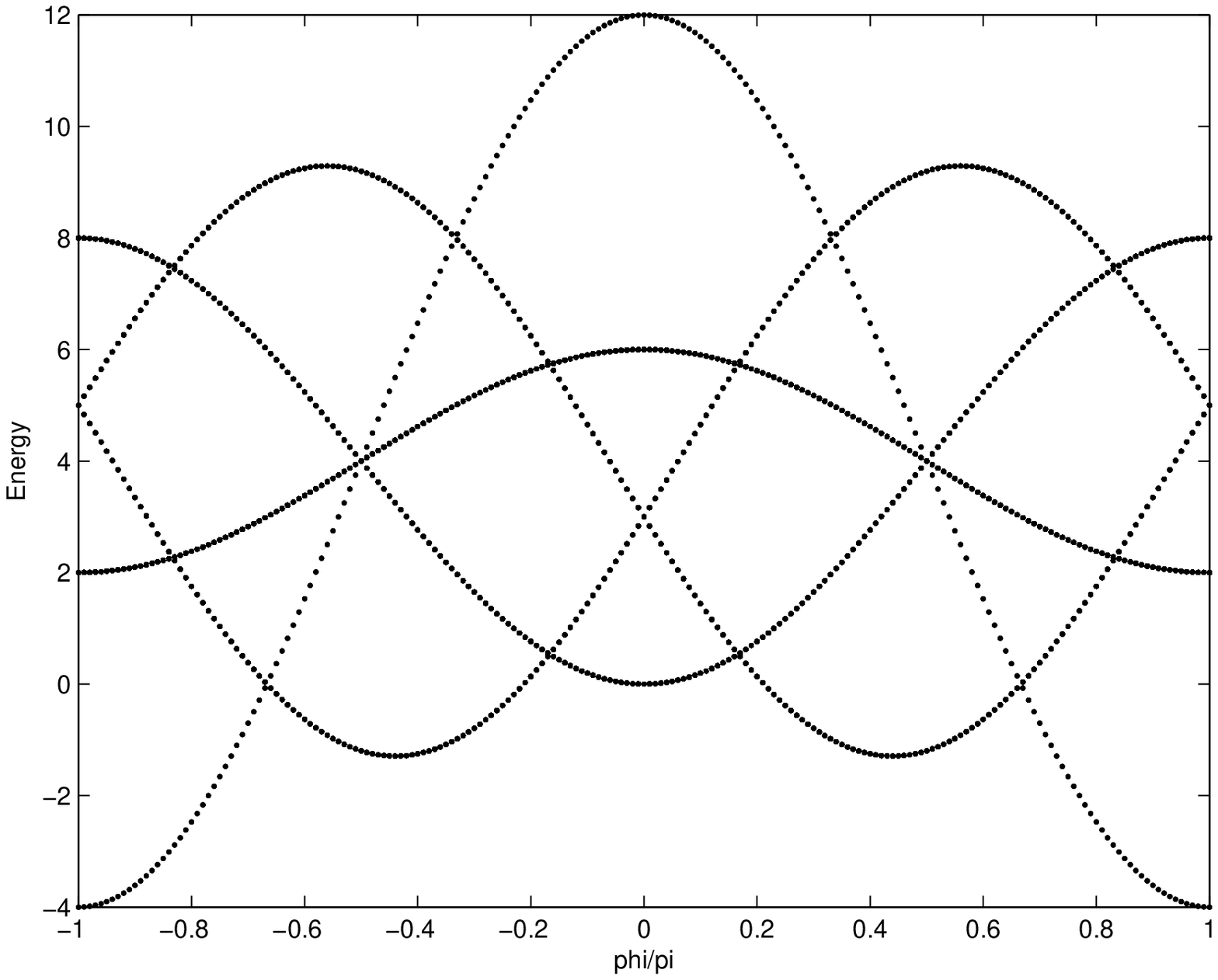}}
\parbox{7cm}{\centering\includegraphics[height=5.5cm]{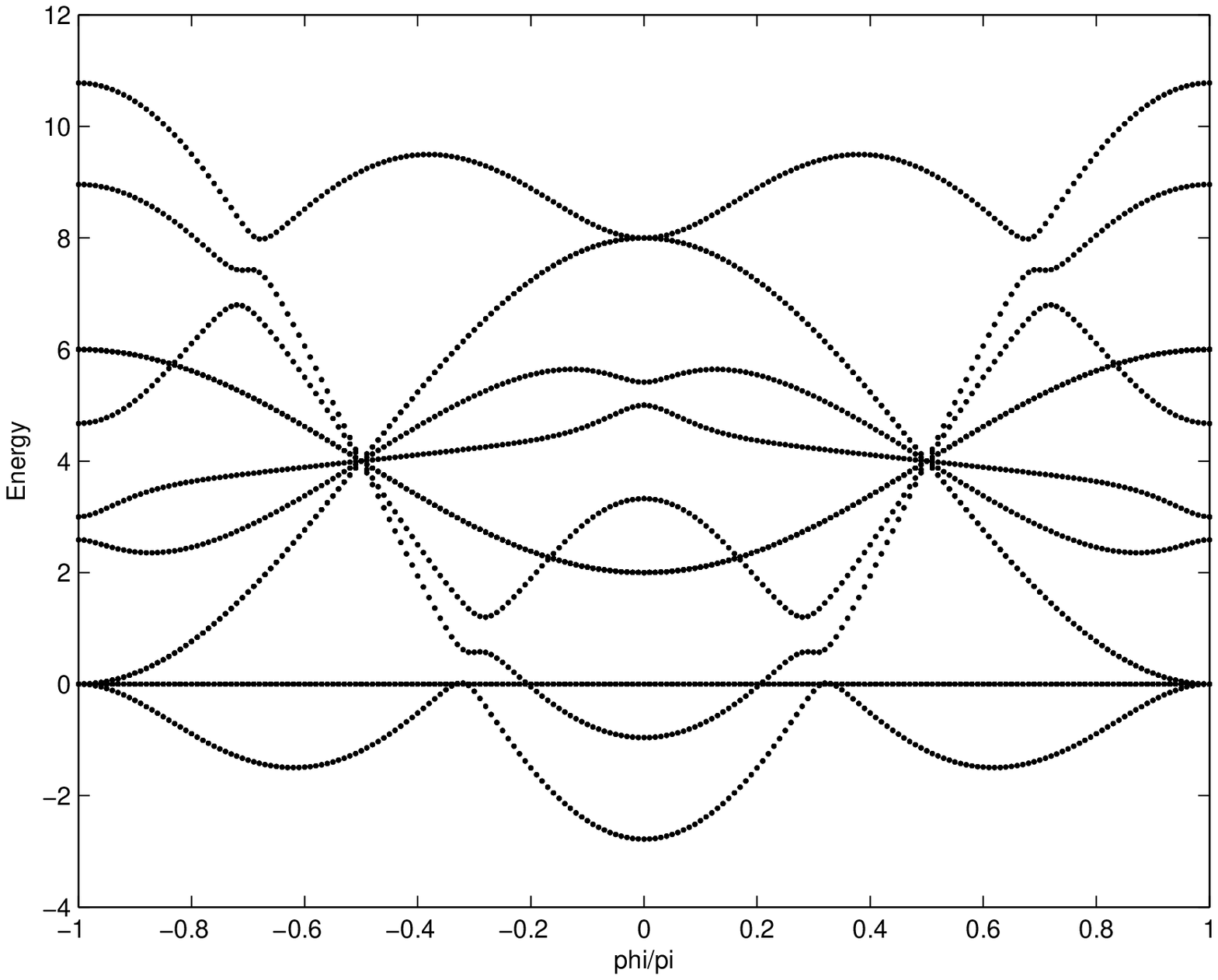}}
\caption{The eigenvalue dependence on the phase for the the Hamiltonian
(\protect\ref{ekv:dil2}).}}
Actually this Hamiltonian can be related with the transformation
(\ref{ekv:transformation}) to a completely diagonal Hamiltonian, such that
it is included in the integrable models mentioned in
\cite{Freyhult:2005ws}.
In figure \ref{fig:spektra70} the graph to the left shows how the
energy eigenvalues of the Hamiltonian (\ref{ekv:dil2}) depends
on the phase $\phi$. The graph to the right shows the eigenvalues,
of the Hamiltonian if we change the sign in front of the second and
 third term in (\ref{ekv:dil2}), depending on the phase $\phi$.
The graph to the right looks very amusing. It looks very similar to
the graph to the left if that is turned upside down and deformed in a
considerable
symmetrical way.

%%%%%%%%%%%%%%%%%%%%%%%%%%%%%%%%%%%%%%%%%%%%%%%%%%%%%%%%%%%%%%%%%%%%%%%%%%
%
%                      Section 6: Broken Z3xZ3
%
%%%%%%%%%%%%%%%%%%%%%%%%%%%%%%%%%%%%%%%%%%%%%%%%%%%%%%%%%%%%%%%%%%%%%%%%%%
\section{Broken $\mathbf{Z_3\times Z_3}$ symmetry}

Relaxing the one-loop finiteness condition (\ref{finite}), by
choosing $h^{000}=h^{222}=0$ and $h^{111}\propto h$ in the superpotential
(\ref{deformed-superpotential}) breaks the $Z_3\times Z_3$ symmetry.
The superpotential is
%%%%
\be \label{ekv:broken}
 W\propto\tr\left[\Phi_0\Phi_1\Phi_2-q\Phi_1\Phi_0\Phi_2 +
\frac{h}{3} \Phi_1^3 \right]\,,
\ee
%%%%%%%%%%
 where an overall factor is
excluded. This superpotential is actually easier to study since the
dilatation operator has homogeneous vacua $|0\rangle|0\rangle\ldots
|0\rangle $ and $|2\rangle|2\rangle\ldots |2\rangle $. The
mixing-matrix for the anomalous dimensions has the form of a
spin-chain Hamiltonian arising from R-matrices found by
Fateev-Zamolodchikov (or XXZ) \cite{Zamolodchikov:1980ku} and the
Izergin-Korepin \cite{Izergin:1980pe}. This type of models were
considered in \cite{Alcaraz:2003ya} even though the authors never
completely classified them. They have a $U(1)$-symmetry which can be
used to get rid of the phase in the complex variable $h$.

In this setting, there is no longer a cancelation between the
fermion loop and the scalar self-energy. The additional contribution
to the Hamiltonian is of the form (\ref{self-energy-contribution})
(see Appendix \ref{app} for details). The spin chain obtained from
the superpotential \ref{ekv:broken} is, with $q=-1$,
\be
H=\left(\begin{array}{ccc|ccc|ccc}
0 & & & & & & & &  \\
 & 1-\frac{h^*h}{2} & & 1 & & & & & \\
& & 1 & & h^* & & 1 & & \\
\hline & 1 & & 1-\frac{h^*h}{2} & & & & & \\
& & h & & 0 & & h & & \\
& & & & & 1-\frac{h^*h}{2} & & 1 &  \\
\hline & & 1 & & h^* & & 1 & & \\
& & & & &1 & & 1-\frac{h^*h}{2} & \\
& & & & & & & & 0
\end{array} \right) \,.
\ee
 The term $h^*h/2$, is the fermion loop contribution from the self-energy.
 We will show that for the special values $q=-1$ and
$h=e^{i\phi}\sqrt{2}$, this Hamiltonian can be obtained from the
spin-1 XXZ R-matrix. The phase of $h$ is redundant, the energy does
not depend on it, and can be phased away through the transformation
$|\tilde{1}\rangle=e^{-i\phi/2} |1\rangle$. The R-matrix for the
XXZ-model is \cite{Zamolodchikov:1980ku}
%%%%
\be\label{XXZ}
R(u)=\left(\begin{array}{ccc|ccc|ccc}
\;s & & & & & & & &  \\
 & \;t & & \;r & & & & & \\
& & \;T & & \;a^* & & \;R & & \\
\hline &\; r\; & & \;t & & & & & \\
& & \;a & & \;\s & & a & & \\
& & & & & \;t & & \;r &  \\
\hline & & \;R & & \;a^* & &\; T & & \\
& & & & &\;r & & \;t & \\
& & & & & & & & \;s
\end{array} \right)
\;\;\; \begin{array}{l}
s=1 \\
t=\epsilon J\sinh(u)\\
r=J\sinh 2\eta \\
a=e^{i\phi} J \frac{\sinh u \sinh 2\eta}{\sinh(u+\eta)}\\
R=J\frac{\sinh \eta \sinh 2\eta }{\sinh (u+\eta)} \\
T=J\frac{\sinh u \sinh (u-\eta)}{\sinh(u+\eta)} \\
\sigma= \epsilon t+R \\
J=\frac1{\sinh (u+2\eta)}
\end{array}
\ee
%%%%%%%%%%
where $\epsilon=\pm 1$. The $\epsilon$ in $t$ in (\ref{XXZ}) is
 added after checking that the R-matrix still satisfies
Yang-Baxter equation.
 If we put $u=0$, the R-matrix becomes the permutation matrix. Thus, a
Hamiltonian can be obtained from the R-matrix by the usual procedure
$H=P R'|_{u=0}$. Performing the derivatives at the point $u=0$ gives
\bea s' &=&0\,, \qquad t'=\epsilon \frac1{\sinh 2\eta}\,,\qquad
r'=-\frac{\cosh
2\eta}{\sinh 2\eta}\,, \qquad a'=e^{i\phi} \frac1{\sinh \eta}\,,\qquad \nonumber \\
R'&=&-\frac{\cosh \eta}{\sinh \eta} -\frac{\cosh 2\eta}{\sinh
2\eta}\,\,\,\,\,, \qquad T'=-\frac1{\sinh 2\eta}\,,\qquad \sigma'=
\epsilon t'+R'
 \,.\eea Multiplying all parameters  with $\sinh 2\eta$, the new variables,
evaluated at $\eta=\pi/4 $, leads to \be \tilde{s}'=0 \,,
\hspace{0.3cm} \tilde{t}'=-1 \,,\hspace{0.3cm} \tilde{r}'=0 \,,
\hspace{0.3cm} \tilde{a}'=e^{i\phi}
\sqrt{2}\,,\hspace{0.3cm}\tilde{R}'=-1 \,, \hspace{0.3cm}
\tilde{T}'=-1 \,, \hspace{0.3cm}\tilde{\sigma}'= 0 \,. \ee with the
corresponding Hamiltonian
%%%
\be H=\left(\begin{array}{ccc|ccc|ccc}
0 & & & & & & & &  \\
 & 0 & & \pm 1 & & & & & \\
& & -1 & & e^{-i\phi}\sqrt{2} & & -1 & & \\
\hline & \pm 1 & & 0 & & & & & \\
& & e^{i\phi}\sqrt{2} & & 0 & & e^{i\phi}\sqrt{2} & & \\
& & & & & 0 & & \pm 1 &  \\
\hline & & -1 & & e^{-i\phi}\sqrt{2} & & -1 & & \\
& & & & &\pm 1 & & 0 &  \\
& & & & & & & & 0
\end{array} \right)
\ee
%%%%
 If we make the choice $\epsilon=- 1$, this is the spin chain Hamiltonian
with deformation $h=e^{i\phi}\sqrt{2}$ and $q=-1$ ! Looking at the
left graph of Figure \ref{fig:spektra2} of a four-site spin chain we
see that two lines cross at this point.
 \FIGURE[t]{\label{fig:spektra2}
\parbox{7cm}{\centering\includegraphics[height=5.5cm]{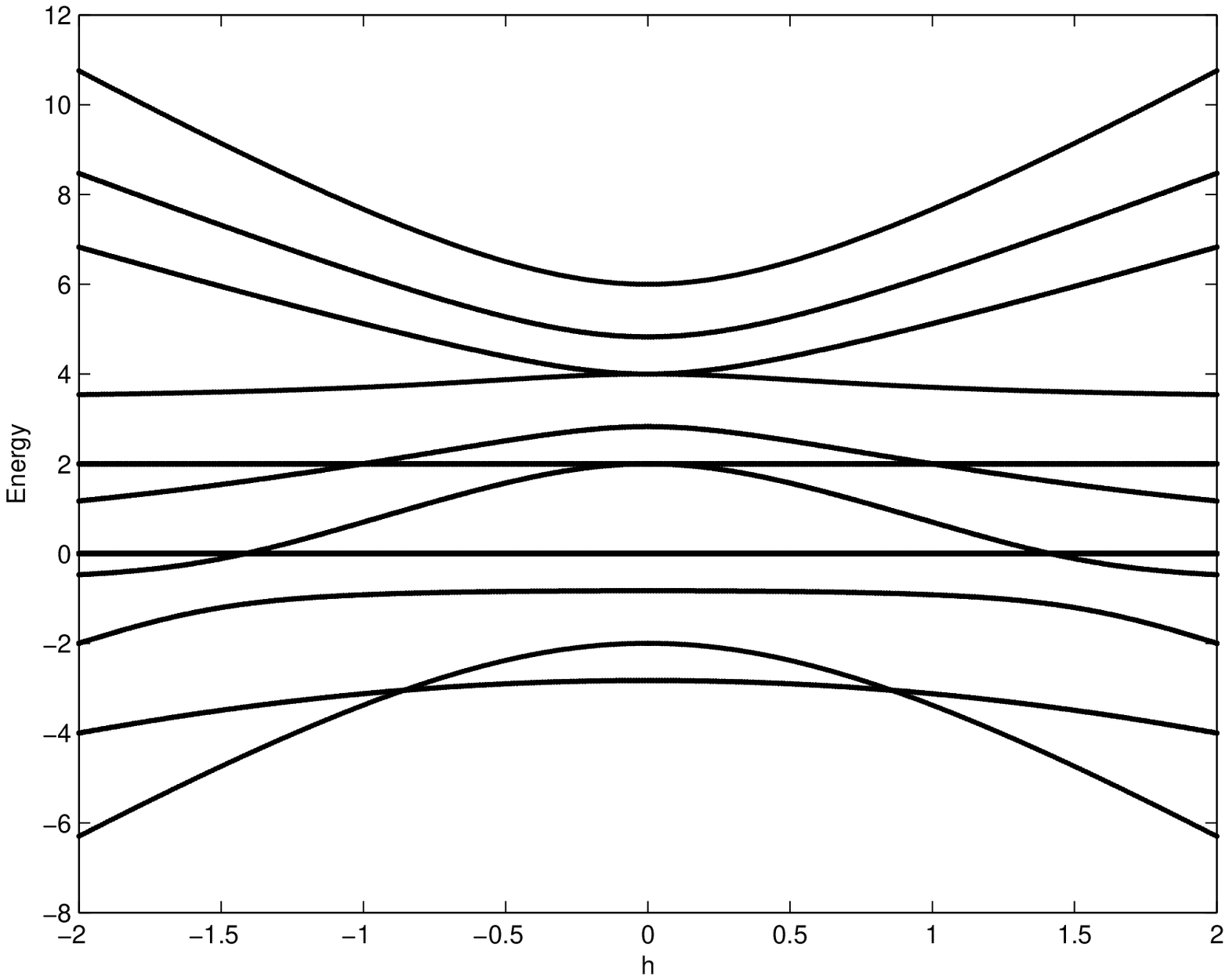}}
\parbox{7cm}{\centering\includegraphics[height=5.5cm]{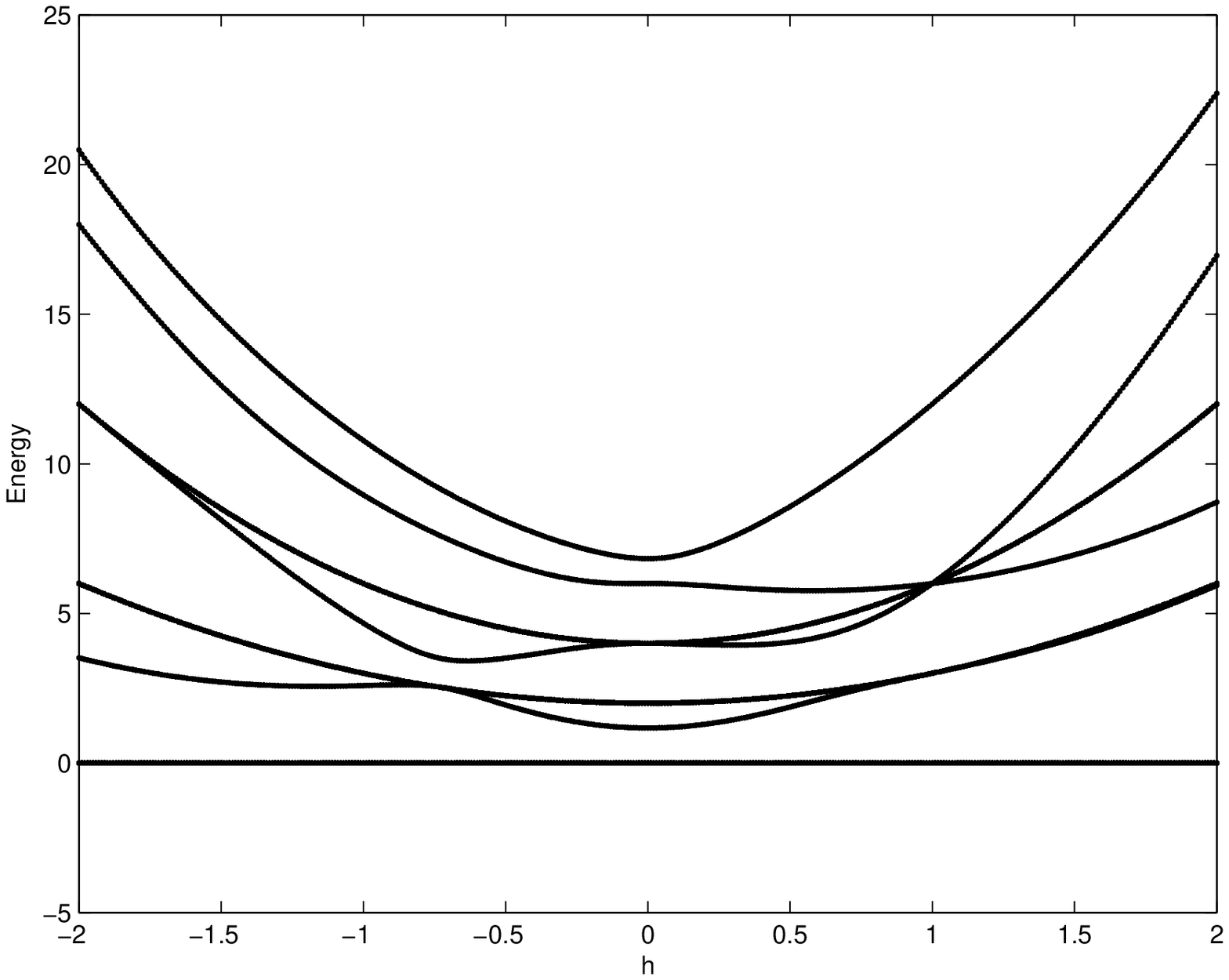}}
\caption{To the left is the spectra for the case $h^{000}=h^{222}=0$
and $h^{111}=h$ depending on $h$  when
 $q=-1$, and to the right is the spectra for the case $h^{III}$ all equal
to $h$ (up to a constant factor) depending on $h$ when $q=-1$}} This
might, however, just be a coincidence. A special feature with $q=-1$
is that there is a $Z_2$-symmetry due to the invariance under
exchange of the fields $\Phi_0$ and $\Phi_2$.

The right graph shows the same spectrum, but with all couplings
$h^{III}$ equal to $h$, up to a constant factor. The point
$h=1-\sqrt{3}$ is special, since at this point the transformation
(\ref{ekv:transformation}) is ``self-dual" , which means here that
$\tilde{q}=q$ and $\tilde{h}=h$.

%%%%%%%%%%%%%%%%%%%%%%%%%%%%%%%%%%%%%%%%%%%%%%%%%%%%%%%%%%%%%%%%%%%%%%%%%%
%
%                      Conclusions
%
%%%%%%%%%%%%%%%%%%%%%%%%%%%%%%%%%%%%%%%%%%%%%%%%%%%%%%%%%%%%%%%%%%%%%%%%%%

\section{Conclusions}
We have studied the dilation operator, corresponding to the general
Leigh-Strassler deformation with $h$ non-zero of $\mathcal{N}=4$
SYM, in order to find new integrable points in the parameter-space
of couplings. In particular we have found a relationship between the
$\gamma$-deformed SYM and a site dependent spin-chain Hamiltonian.
 When all
parameters $\g_i$ are equal, this relates an entirely $q$-deformed to an
entirely $h$-deformed superpotential. For $q=0$ and the  $h=e^{i\theta}$,
where $\theta$ is real, we have
found a new R-matrix (see \ref{Rmatrix1}).

We found a way of representing a general ansatz for the R-matrix,
with the right form to give the dilatation operator, which makes the
structure of the Yang-Baxter equations clear. The equations can be
represented in terms of rectangular objects, which reveals that the
underlying structure is a generalized version of the structure of
the eight-vertex model. We presented all values of the parameters
$q$ and $h$ for which the spin-chain Hamiltonian can be obtained
from R-matrices with a linear dependence on the spectral parameter.
Most of them were related to the $q$-deformed case through a simple
shift of basis  with a real phase $\beta$, or a shift with a twist
with the phase $\pm 2\pi/3$, which reflects the $Z_3$-symmetry.

We also found a new hyperbolic R-matrix (\ref{ekv:dil2}) which,
through a simple change of basis, gives a Hamiltonian with only
diagonal terms which was included in the cases studied in
\cite{Freyhult:2005ws}. We had a brief look at a case with broken
$Z_3\times Z_3$ symmetry and found that the matrix of anomalous
dimensions can for some special values of the parameters be obtained
from the Fateev-Zamolodchikov R-matrix.

We conjecture that the Yang-Baxter equations found for the general
R-matrix have a solution which is a generalized version of the
solution to the eight-vertex model. If this solution exists, it is
plausible that there will exist more points in the parameter space
for which the dilatation operator is integrable. To find a general
solution to these equations would be of interest in its own right.
{}From a mathematical point of view, it is then interesting to
generalize the solution to an R-matrix of arbitrary dimension.

The found relationship between the q- and the h-deformed
superpotential should be visible in the dual string theory, and
 should also give a clue of what that string theory looks like. Another
 way to approach the problem, as mentioned in
\cite{Frolov:2005iq}, is to first find a coherent state spin chain
and from that reconstruct the dual geometry. The coherent state spin
chain \cite{Frolov:2005iq} is valid for small $\beta$, \textit{i.e.}
$q\approx 1$. We believe that making use of the basis transformation
(\ref{ekv:transformation}) makes it possible to create a coherent
state spin chain for $q\approx 1$ and small $h$ acting with the
transformation (\ref{ekv:transformation}) on a $q$ close to one
gives a new $q$ close to one and a new small $h$. We also hope that
due to the relation between vanishing $h$ and vanishing $q$ it is
possible to write a coherent sigma model for both $q$ and $h$ close
to one. It would then be very interesting to find the dual geometry,
which corresponds to a further away deformation of the
$\mathcal{N}=4$ SYM.

One other thing of interest is to extend the analysis to other
sectors of the theory and to higher loop order. In the
$\beta$-deformed case it is possible to argue that the integrability
holds to higher loop order \cite{Frolov:2005ty}, because the
dilatation operator is related with a unitary transformation to the
case of the usual $\mathcal{N}=4$ SYM. In the same way can we argue
about the $h$-deformed case, even though we have to consider the
induced effects of the spin chain periodicity.

\acknowledgments
We would like to thank Lisa Freyhult,
 Charlotte Kristjansen, Sergey Frolov, Anna Tollst\'{e}n, Johan Bijnens  and
Matthias Staudacher for interesting discussions and commenting the
 manuscript. We would also like to thank Anna Tollst\'{e}n for her contribution
to the solution of the linear ansatz.

%%%%%%%%%%%%%%%%%%%%%%%%%%%%%%%%%%%%%%%%%%%%%%%%%%%%%%%%%%%%%%%%%%%%%%%%%%
%
%                      Appendix
%
%%%%%%%%%%%%%%%%%%%%%%%%%%%%%%%%%%%%%%%%%%%%%%%%%%%%%%%%%%%%%%%%%%%%%%%%%%

\newpage
\appendix

\section{Yang-Baxter equations for the general case}\label{app1}
The functions in the R-Matrix (\ref{ekv:rmatrix}) are expressed in
terms of the functions $\o_i$, $\bar{\o}_i$ and $\gamma_{i}$~as
\bea\label{parameter}
a(u)&=&\g_1(u)+\g_2(u)+\g_3(u)\,,\nonumber\\
b(u)&=&\g_1(u) e^{i2\pi/3}+\g_2(u)e^{-i2\pi/3}+\g_3(u)\,,\nonumber \\
\bar{b}(u)&=&\g_2(u) e^{i2\pi/3}+\g_1(u)e^{-i2\pi/3}+\g_3(u)\nonumber \\
c(u)&=&\o_1(u)+\o_2(u)+\o_3(u)\,,\nonumber\\
c(u)&=&\bar{\o}_1(u)+\bar{\o}_2(u)+\bar{\o}_3(u)\,,\\
d(u)&=&\o_2(u) e^{i2\pi/3}+\o_1(u)e^{-i2\pi/3}+\o_3(u)\,, \nonumber\\
\bar{d}(u)&=&\bar{\o}_1(u) e^{i2\pi/3}+\bar{\o}_2(u)
e^{-i2\pi/3}+\bar{\o}_3(u)\,, \nonumber \\
e(u)&=&\o_1(u) e^{i2\pi/3}+\o_2(u)e^{-i2\pi/3}+\o_3(u)\,, \nonumber \\
\bar{e}(u)&=&\bar{\o}_2(u)
e^{i2\pi/3}+\bar{\o}_1(u)e^{-i2\pi/3}+\bar{\o}_3(u)\,, \nonumber \, .
\eea
Yang-Baxter equations from the R-matrix ansatz (\ref{ekv:Rmatrix2})
read
 \bea \label{ekv:YB}
\o_{n+1}\o_{n+2}'\g_3''-\o_{n+2}\o_{n+1}'\g_2''
+\g_3\bar{\o}_{n}'\bar{\o}_{n+1}''-\g_2\bar{\o}_{n+1}'\bar{\o}_{n}''
+\bar{\o}_{n+1}\g_2'\o_{n+1}''-\bar{\o}_{n}\g_3'\o_{n+2}''
&=&0 \,,\nonumber \\
\o_{n+1}\o_{n+2}'\g_1''-\o_{n+2}\o_{n+1}'\g_3''
+\g_1\bar{\o}_{n+2}'\bar{\o}_{n}''-\g_3\bar{\o}_{n}\bar{\o}_{n+2}''
+\bar{\o}_{n}\g_3'\o_{n+1}''-\bar{\o}_{n+2}\g_1'\o_{n+2}''
&=&0 \,,\nonumber \\
\o_{n+1}\o_{n+2}'\g_2''-\o_{n+2}\o_{n+1}'\g_1''
+\g_2\bar{\o}_{n+1}'\bar{\o}_{n+2}''-\g_1\bar{\o}_{n+2}\bar{\o}_{n+1}''
+\bar{\o}_{n+2}\g_1'{\o}_{n+1}''-\bar{\o}_{n+1}\g_2'{\o}_{n+2}''
&=&0\,, \nonumber \\
\o_1\bar{\o}_{n+1}'\o_{2}''-\bar{\o}_1\o_{2n+1}'\bar{\o}_{3}''+
\o_2\bar{\o}_{n+2}'\o_{0}''-\bar{\o}_2\o_{2n-1}'\bar{\o}_{1}''+
\o_0\bar{\o}_{n}'\o_{1}''-\bar{\o}_0\o_{2n}'\bar{\o}_{2}''
&=&0 \,,\nonumber \\
\g_2\o_{n+1}'\g_1''+\g_3\o_{n-1}'\g_2''+\g_1\omega_n'\g_3''
-\o_1\g_{n}'\o_2''-\o_2\g_{n+1}'\o_3''-
\o_3\g_{n-1}'\o_1''
&=&0 \,,\nonumber\\
\g_1\bar{\o}_{n-1}'\g_2''+\g_2\bar{\o}_{n+1}'\g_3''+\g_3\bar{\omega}_n'\g_1''
-\bar{\o}_1\g_{n-1}'\bar{\o}_2''-\bar{\o}_2\g_{n+1}'\bar{\o}_3''-
\bar{\o}_3\g_n'\bar{\o}_1''
&=&0 \,,\nonumber\\
\bar{\o}_{n+1}\bar{\o}_{n+2}'\o_{n+1}''-\bar{\o}_{n+2}\bar{\o}_{n+1}'\o_{n}''
-\o_n\g_3'\g_1''+\o_{n+1}\g_1'\g_3''
-\g_1\o_{n+1}'\bar{\o}_{n+2}''+\g_3\o_{n}'\bar{\o}_{n+1}''
&=&0 \,,\nonumber\\
\bar{\o}_{n+2}\bar{\o}_{n+1}'\o_{n+2}''-\bar{\o}_{n+1}\bar{\o}_{n+2}'\o_{n}''
-\o_n\g_3'\g_2''+\o_{n+2}\g_2'\g_3''
-\g_2\o_{n+2}'\bar{\o}_{n+1}''+\g_3\o_{n}'\bar{\o}_{n+2}''
&=&0 \,,\nonumber\\
\bar{\o}_{n+1}\bar{\o}_{n+2}'\o_{n+2}''-\bar{\o}_{n+2}\bar{\o}_{n+1}'\o_{n+1}''
-\o_{n+1}\g_1'\g_2''+\o_{n+2}\g_2'\g_1''
+\g_1\o_{n+1}'\bar{\o}_{n+1}''-\g_2\o_{n+2}'\bar{\o}_{n+2}''
&=&0\,,
 \nonumber\\
{\o}_{n+2}{\o}_{n}'\bar{\o}_{n+1}''-\o_{n}\o_{n+2}'\bar{\o}_{n}''
+\bar{\o}_{n+1}\g_3'\g_1''-\bar{\o}_n\g_1'\g_3''
+\g_1\bar{\o}_{n}'\o_{n+2}''-\g_3\bar{\o}_{n+1}'\o_{n}''
&=&0 \,,\nonumber\\
\o_{n}\o_{n+1}'\bar{\o}_{n}''-{\o}_{n+1}{\o}_{n}'\bar{\o}_{n+2}''
-\bar{\o}_{n+2}\g_3'\g_2''+\bar{\o}_{n}\g_2'\g_3''
-\g_2\bar{\o}_{n}'\o_{n+1}''+\g_3\bar{\o}_{n+2}'{\o}_{n}''
&=&0 \,,\nonumber\\
\o_{n+2}\o_{n+1}'\bar{\o}_{n+1}''-\o_{n+1}\o_{n+2}'\bar{\o}_{n+2}''
+\bar{\o}_{n+1}\g_2'\g_1''-\bar{\o}_{n+2}\g_1'\g_2''
+\g_1\bar{\o}_{n+2}'\o_{n+2}''-\g_2\bar{\o}_{n+1}'\o_{n+1}''
&=&0 \,,\nonumber\\
\eea Here, we have defined $\o=\o(u-v)$, $\o'=\o(u)$ and
$\o''=\o(v)$.

\section{Self-energy with broken $\mathbf{Z_3\times Z_3}$ symmetry}\label{app}

We will follow the prescription of \cite{Beisert:2002bb} to compute
the contribution to the Hamiltonian from the superpotential
(\ref{ekv:broken}), when conformal invariance is broken. The
additional terms are coming from the self-energy fermion loop.

The scalar self-energy of the vertices is, in $\mathcal{N}=4$ SYM,
\be \frac{g_{YM}^2(L+1)}{8\pi^2}N:\tr
\left(\bar{\phi}_i \phi_i\right): \,, \ee where
$L=\mbox{log}x^{-2}-\left(1/\epsilon+\g +\mbox{log}\pi+2\right)$.
The scalar-vector contribution to this is
$-\frac{g_{YM}^2(L+1)}{8\pi^2}$, and the fermion loop contribution
is $\frac{g_{YM}^2(L+1)}{4\pi^2}$. Half of the fermion contribution
comes from the superpotential; this is the part which will be
altered by the extra $h$-dependent part of the superpotential.
Hence, the additional term to the new spin chain, besides the F-term
scalar part, is \be \frac{h^*h}{1+q^*q}\frac{g^2(L+1)}{8\pi^2}N :\tr
\left(\bar{\phi}_1\phi_1\right):\,. \ee Then, we will have an
effective scalar interaction which just comes from the F-term (since
we have the  same cancelation as in the $\mathcal{N}=4$
SYM)\cite{Beisert:2002bb} 
%%%%%
\be \label{effective-scalar}
\pm\sqrt{\frac{2}{(1+q^*q)}}\frac{g_{YM}^2 L}{16\pi^2}:V_F: \,,\ee
where \bea V_F &=&\left(\tr\left[\phi_i
\phi_{i+1}\bar{\phi}_{i+1}\bar{\phi}_i
 -q\phi_{i+1}\phi_i\bar{\phi}_{i+1}\bar{\phi}_i
 - q^{*} \phi_i\phi_{i+1}\bar{\phi}_i\bar{\phi}_{i+1} \right]\nonumber\right. \\
&+&\tr\left[ qq^{*}\phi_{i+1}\phi_i\bar{\phi}_i\bar{\phi}_{i+1}
-qh^{*} \phi_{0}\phi_2 \bar{\phi}_{1}\bar{\phi}_{1}
-q^{*}h \phi_{1}\phi_{1}\bar{\phi_2}\bar{\phi}_{0}\right]\nonumber \\
&+&\left.\tr\left[ h\phi_{1}\phi_{1}\bar{\phi}_{0}\bar{\phi}_2
+h^{*} \phi_2\phi_{0}\bar{\phi}_{1}\bar{\phi}_{1} +hh^{*}
\phi_{1}\phi_{1}\bar{\phi}_1\bar{\phi}_1 \right]\right)\,. \eea The
plus-minus sign in (\ref{effective-scalar}) depends on which sign we
choose for the superpotential. Since all terms are multiplied by the
same divergent factor we can set $L=\mbox{log}x^{-2}$, just as in
the case of $\mathcal{N}=4$.
 The contribution from the self-energy to the dilatation operator is
\be \label{self-energy-contribution}
\frac{h^*h}{1+q^*q}(E_{11}\otimes I+I\otimes E_{11}) \,,\ee and the
F-term scalar interaction contribute with \bea
&&\pm\sqrt{\frac{2}{(1+q^*q)}}\left(
 E_{i,i}^{l} E_{i+1,i+1}^{l+1}  -q E_{i+1,i}^{l}
E_{i,i+1}^{l+1}  -q^{*}
E_{i,i+1}^{l} E_{i+1,i}^{l+1} \nonumber \right. \\
&+&qq^{*} E_{i+1,i+1}^{l}   E_{i,i}^{l+1}  -qh^{*}
 E_{i+1,i+2}^{l} E_{i,i+2}^{l+1}   -q^{*}h
E_{1,0}^{l} E_{1,2}^{l+1}  \nonumber
\\&+& \left. hE_{1,2}^{l} E_{1,0}^{l+1}
+h^{*}E_{2,1}^{l} E_{0,1}^{l+1} +hh^{*}E_{1,1}^{l}
E_{1,1}^{l+1}\right)\,. \eea We will now consider the case when
$q=-1$. The total dilatation operator simplifies to \be
H=\left(\begin{array}{ccc|ccc|ccc}
0 & & & & & & & &  \\
 & 1-\frac{h^*h}{2} & & 1 & & & & & \\
& & 1 & & -qh^* & & 1 & & \\
\hline & 1 & & 1-\frac{h^*h}{2} & & & & & \\
& & h & & 0 & & h & & \\
& & & & & 1-\frac{h^*h}{2} & & 1 &  \\
\hline & & 1 & & h^* & & 1 & & \\
& & & & &1 & & 1-\frac{h^*h}{2} & \\
& & & & & & & & 0
\end{array} \right) \,.
\ee
Here we have chosen a relative minus sign between the contribution
from the fermion loop and the scalar interaction term.
%\end{appendix}
\bibliographystyle{JHEP}
\bibliography{leighref}

\end{document}